\documentclass[preprintnumbers,
amsmath,amssymb,floatfix,10pt,prd,onecolumn,
superscriptaddress,longbibliography,nofootinbib]{revtex4-2}
\usepackage{bm}
\usepackage{amsfonts}
\usepackage{latexsym}
\usepackage{amsmath}
\usepackage{palatino}
\usepackage{textcomp}
\usepackage{float}
\usepackage{booktabs}
\usepackage{dcolumn}
\usepackage{dsfont}
\usepackage{ragged2e}
\usepackage{epsfig}
\usepackage[dvipsnames]{xcolor}
\usepackage{hyperref}
\hypersetup{           
    colorlinks=true,                
    breaklinks=true,                
    urlcolor= OrangeRed,                
    linkcolor= magenta,                
    bookmarksopen=false,
    filecolor=black,
    citecolor=blue,
    linkbordercolor=red}
\usepackage{graphicx}
\usepackage{orcidlink}
\usepackage[T1]{fontenc}

\begin{document}
\title{Investigating Optical and Ring-Down Gravitational Wave Properties of a Rotating Black Hole in a Dehnen Galactic Dark Matter Halo}

\author{Mrinnoy M. Gohain\orcidlink{0000-0002-1097-2124}}
\email[Email:]{mrinmoygohain19@gmail.com}
\affiliation{%
 Department of Physics, Dibrugarh University, Dibrugarh \\
 Assam, India, 786004}
 
\author{Dhruba Jyoti Gogoi \orcidlink{0000-0002-4776-8506}}%
 \email[Email:]{moloydhruba@yahoo.in}
\affiliation{%
 Department of Physics, Moran College, Moranhat, Charaideo 785670, Assam, India \\
 Assam, India, 786004}
 \affiliation{Theoretical Physics Division, Centre for Atmospheric Studies, Dibrugarh University, Dibrugarh, Assam, India 786004}
 
\author{Kalyan Bhuyan\orcidlink{0000-0002-8896-7691}}%
 \email[Email:]{kalyanbhuyan@dibru.ac.in}
\affiliation{%
 Department of Physics, Dibrugarh University, Dibrugarh \\
 Assam, India, 786004}%
 \affiliation{Theoretical Physics Division, Centre for Atmospheric Studies, Dibrugarh University, Dibrugarh, Assam, India 786004}

\author{Prabwal Phukon\orcidlink{0000-0002-4465-7974}}%
 \email[Email:]{prabwal@dibru.ac.in}
\affiliation{%
 Department of Physics, Dibrugarh University, Dibrugarh \\
 Assam, India, 786004}%
 \affiliation{Theoretical Physics Division, Centre for Atmospheric Studies, Dibrugarh University, Dibrugarh, Assam, India 786004}



\begin{abstract}
We present a comprehensive study of the optical and dynamical properties of a rotating black hole immersed in a Dehnen-type (1,4,0) galactic dark matter halo, modeled by a double power-law density profile commonly used to describe realistic galactic cores. By extending our previous Schwarzschild–Dehnen solution using a modified Newman–Janis algorithm, we construct a Kerr-like axisymmetric spacetime that smoothly incorporates both black hole rotation and the influence of the surrounding dark matter halo. We systematically investigate the effects of the halo parameters-the central density and halo radius-on horizon structure, the shape and extent of the ergoregion, and the null geodesics associated with black hole shadows. Our results show that the presence of a dense or extended halo expands the event horizon and ergoregion, and significantly alters the size and distortion of the black hole shadow. Furthermore, by applying the WKB approximation to scalar field perturbations, we compute the quasinormal mode (QNM) spectra and demonstrate that the frequencies and damping times of ringdown signals are highly sensitive to the halo profile. These results open promising avenues for probing the dark matter environment of astrophysical black holes through black hole imaging and gravitational wave observations.
\end{abstract}

\maketitle

\section{Introduction}
The idea that black holes (BHs) are entirely isolated in the universe is not very obvious.  They may be found in environments that can complicated and dynamic. Specifically, there is compelling evidence that active galactic nuclei are powered by supermassive BHs \cite{Rees1984Sep,Kormendy1995Sep}.  Additionally, there is also strong evidence that most galaxies are surrounded by a halo of dark matter \cite{Bertone2018Oct}.  For particular kinds of dark-matter halo patterns, a number of significant works based on BH-dark matter composite systems have been investigated.  For example, in a Hernquist-like dark matter profile, Xavier et al. investigated the shadows of non-rotating BH \cite{Xavier2023Mar}. The impact of the Hernquist dark matter halo profile on gravitational wave propagation and geodesics was examined by Cardoso et al. \cite{Cardoso2022Mar}.  A new BH solution surrounded by a dark matter halo at the galactic center of the M87 galaxy has been suggested by Jusufi et al. \cite{Jusufi2019Aug} based on the universal rotation curve (URC) dark matter profile.  A spherical model with a Schwarzschild BH and a piecewise distribution of dark matter surrounding it was examined by Konoplya \cite{Konoplya2019Aug}.  Zhidenko and Konoplya \cite{Konoplya2022Jul} derived BH solutions in the presence of different dark matter halo characteristics in another study. Rotating BH at the center of the Sgr* galaxy with cold dark matter and scalar field dark matter halos was investigated by Hou et al. \cite{Hou2018Jul}.  The optical properties of a rotating BH-dark matter system with a pseudo-isothermal dark matter halo profile were examined by Yang et al. \cite{Yang2024Jan}.  The thermodynamic properties of a BH submerged in a perfect fluid dark matter halo were investigated by Liang et al. \cite{Liang2023Nov}.  Carvalho \cite{Carvalho2023Dec} investigated the Einstein-Gauss-Bonnet (EGB) BH thermodynamics encircled by three distinct dark matter halo distributions. In a study of Kerr BHs enveloped by perfect fluid dark matter, Anjum et al. \cite{Anjum2023May} examined the photon orbits, naked singularities, and shadows in relation to the Event Horizon telescope's observations.  In the context of the Bumblebee Gravity, Cappozziello et al. \cite{Capozziello2023May} investigated the consequences of a dark matter spike close to the supermassive BH M87.  Jusufi \cite{Jusufi2023Feb} introduced a spherically symmetric and asymptotically flat BH solution in which Einstein clusters develop the dark matter halo and dark matter is composed of weakly interacting particles orbiting a supermassive BH in the galactic center. Pantig and \"{O}v\"{g}un \cite{Pantig2022May}, investigated the influence of different dark matter  halo profiles on the weak deflection angle. Some other interesting works related to BH systems in dark matter environments can be found in Refs. \cite{Stuchlik2021Nov,Pantig2023Jan,Ovgun2024Apr}.

The Dehnen density profile \cite{Dehnen1993Nov,Mo2010May} is commonly studied while dealing with dwarf galaxies, which often do not host BHs at their centers. However, recent observations suggest that massive BHs may also be present in these dwarf galaxies. Specifically, it was reported that a supermassive BH (SMBH) with a mass of around $2.00 \times 10^5 M_{\odot}$ resides in the dwarf galaxy Mrk 462 \cite{BibEntry2024Jun}. Additionally, a study on BH-triggered star formation in the dwarf galaxy Henize 2-10 identified an SMBH with a mass of around $1.00 \times 10^6 M_{\odot}$. Furthermore, a dynamical study of dark matter using photometric and spectroscopic data revealed the presence of a BH in Leo I, with a mass of $3.3 \pm 2 \times 10^6 M_{\odot}$, accounting for 13\% of the total mass of the Leo I galaxy \cite{Bustamante-Rosell2021Nov}.

The study of rotating BHs surrounded by dark matter has become a popular field of study, connecting theoretical astrophysics and observational cosmology. Dark matter, an invisible constituent of the universe, is important for galaxy formation and evolution, and usually believed to exist in vast halos within which galaxies reside \cite{Mollicone2025Jan}. BHs, especially supermassive BHs located at galactic centers, gravitationally interact with these dark matter halos, resulting in interesting astrophysical phenomena \cite{Biswas2019Sep}. Advanced observation recently, using ultra-high angular resolution images such as supermassive BH M87* \cite{TheEventHorizonTelescopeCollaboration2019Apr} and Sagittarius A* \cite{EventHorizonTelescopeCollaboration2022May}, have revealed dark central space with bright surrounding rings, both corresponding to the BH's shadow and the photon ring and providing useful insights into the geometry of spacetime around such entities. Such tests present a benchmark for numerous theories, including the role of dark matter - BH composite systems \cite{Jusufi2019Aug}.
Theoretically, dark matter surrounding BHs may alter the spacetime geometry, which has implications for properties like shadow of the BH, quasinormal modes, and accretion disk physics. Modeling this interaction can be addressed through methods involving the construction of new BH solutions with dark matter density profiles included.

{ BHs not only play a central role in modern astrophysics but also serve as natural laboratories for testing the predictions of general relativity in the strong-field regime. A landmark achievement in this context was the direct detection of gravitational waves (GWs) by the LIGO Scientific Collaboration on September 14th, 2015 \cite{LIGOScientific:2016aoc}. This breakthrough provided an unprecedented observational window into the dynamics of compact objects, confirming that GWs are generated by the acceleration of massive bodies and propagate as ripples through spacetime, precisely as predicted by Einstein’s field equations. When two BHs merge, the remnant object emits GWs in the form of a distinctive ringdown signal, which can be described in terms of quasinormal modes (QNMs). QNMs reflect the characteristic oscillations of the resulting BH, with each mode defined by a complex frequency whose real part denotes the oscillation frequency and whose imaginary part governs the damping rate. These modes are called "quasinormal" because, unlike pure normal modes, they decay over time due to energy lost via gravitational radiation \cite{Vishveshwara:1970zz, Press:1971wr, Kokkotas:1999bd}. The analysis of QNMs in gravitational wave data enables the determination of the mass and spin of the final BH, offers deep insights into spacetime structure near the event horizon, and allows for experimental tests of gravitational physics well beyond the weak-field approximation \cite{Li:2021zct}. Recent years have seen a surge of interest in studying QNMs within various modified gravity frameworks and alternative BH models \cite{Anacleto:2021qoe, Lambiase:2023hng}. In this context, exploring how environmental effects-such as the presence of a dark matter halo-influence the QNM spectrum is of prime importance for interpreting gravitational wave signals and for unveiling the fundamental properties of BHs.
}

In this paper, we want to study the optical properties of a rotating analogue of a BH immersed in a Dehnen dark matter halo. To achieve this, we begin by applying the modified Newman Janis algorithm on our recently derived non-rotating version of the Dehnen dark matter-BH composite system and obtained the rotating analogue of the system. In Section \ref{sec2}, we review our previously derived Dehnen dark matter-BH system. In Section \ref{sec3}, we apply the modified Newman-Janis algorithm to obtain the rotating BH solution. In Section \ref{sec4}, we obtain the geodesic equations through Hamilton Jacobi formalism and thereafter derived the shadows as well as distortion of the BH. In Section \ref{energyemission}, we investigate the energy emission rate of the BH. In Section \ref{sec5}, we study the QNMs associated with the BH spacetime. Finally, in Section \ref{conc} we summarize our results of the work.

\section{Dark Matter Density Profile and Metric Function}
\label{sec2}
Recently, we  derived the solution of a non-rotating and uncharged BH solution in the vicinity of a Dehnen-type dark matter halo \cite{Gohain2024Dec}. Let us now briefly discuss the theoretical framework of the BH solution in the presence of the Dehnen dark matter. The density profile of Dehnen dark matter halo is a special case of a double power-law profile given by \cite{Mo2010May}
\begin{equation}
\rho_{dm} = \rho_s \left(\frac{r}{r_s}\right)^{-\gamma } \left[\left(\frac{r}{r_s}\right)^{\alpha }+1\right]^{\frac{\gamma -\beta }{\alpha }}
\label{doub_pow_law}
\end{equation}
In the Dehnen profiles, where $\gamma$ determines the specific variant of the profile. Some of the allowed variants of the Dehnen profile can be obtained from setting $(\alpha,\beta,\gamma) = (1, 4, \gamma)$. The values of $\gamma$ lies within $[0,3]$. For instance, $\gamma = 3/2$ has been used to fit the surface brightness profiles of elliptical galaxies which closely resembles the de Vaucouleurs $r^{1/4}$ profile \cite{Shakeshaft1974Jan}. Recently, Pantig and \"{O}v\"{g}un \cite{Pantig2022Aug} studied the effect of Dehnen dark matter halo in an ultrafaint dwarf galaxy. In our earlier work we derived the solution for the parameters $(\alpha, \beta, \gamma) = (1, 4, 0)$. This gave:
\begin{equation}
\rho_{D} = \frac{\rho_s}{\left(\frac{r}{r_s}+1\right)^4}
\label{dehnen_profile}
\end{equation}
where $\rho_s$ and $r_s$ denote the central halo density and the halo core radius respectively. 
In a recent work, we derived an unique non-rotating uncharged BH solution in the presence of a Dehnen dark matter halo \cite{Gohain2024Dec}, where we implemented the method earlier used by Jusufi et al \cite{Jusufi2019Aug} to produce an effective metric in the presence of a dark matter halo. They derived a Schwarzschild BH by including the correction terms arising due to the presence of a URC type dark matter halo. They further used the Newman-Janis algorithm to produce the rotating version of the obtained BH solution. Inspired by this, in this work we shall use the Newman-Janis method to deduce the rotating analogue of the BH immersed in the Dehnen dark matter halo. In our previous work \cite{Gohain2024Dec}, using the Dehnen profile \eqref{dehnen_profile} and a spherically symmetric metric 
\begin{equation}
	\begin{aligned}
ds^2 &= -f(r) dt^2 + \frac{1}{f(r)}dr^2 + r^2 d\Omega^2; \\& d\Omega^2 = d\theta^2 + \sin \theta d\phi^2,
\end{aligned}
\label{metric}
\end{equation}
we derived the lapse function $f(r)$ in the following form 
with \begin{equation} 
f(r) = 1 - \frac{2M}{r} - \frac{4 \pi  r_s^3 \left(r_s+2 r\right) \rho _s}{3 \left(r_s+r\right)^2}.
\label{fin_metric_funct}
\end{equation}
This is the effective metric function of our composite dark matter-BH system. Clearly, in the limit of $\rho_s \to 0$, one recovers the Schwarzschild metric. The Schwarzschild-Dehnen metric have recently received popularity among researchers in studying various BH properties as can be found in the Refs. \cite{Alloqulov2025Apr,Al-Badawi2024Dec,Al-Badawi2025Jan,Jha2025Mar,Hosseinifar2025Mar}.

\section{Rotating BH in Dehnen Dark Matter Halo}
\label{sec3}

In order to generate rotating BH solutions from non-rotating ones, the Newman-Janis Algorithm (NJA) \cite{Newman1965Jun,Newman1965Jun1} serves as an effective technique.  The method uses a complex coordinate transformation, that is applied to a static, spherically symmetric metric.  The NJA was first used to extract the rotating Kerr BH solution from the non-rotating Schwarzschild BH metric \cite{Drake1997Jul}. Since then, it has been extended to a number of different spacetime types \cite{Brauer2015Jan,Lombardo2004Feb,Kim2025Jan,Abbas2024Apr,Alexeyev2025Mar,Jafarzade2025Jun,Fazzini2025Feb,Li2025Jan,Fathi2025Mar,Zahid2025Feb,Raza2025Jan}.  The fundamental motivation behind this method is the fact that almost all exact BH solutions are stationary due to the symmetry simplifications required for solvability, but the majority of astrophysical BHs rotate rather than stationary.

We shall now proceed to generalize our spherical symmetric non-rotating Schwarzschild-Dehnen BH to a rotating one using the NJA.  To do this, firstly we need to transform the Boyer-Lindquist (BL) coordinates $(t,r,\theta,\phi)$ to the Eddington-Finkelstein (EF) coordinates $(u,r,\theta,\phi)$.  This is done through the following coordinate redefinition\cite{Azreg-Ainou2014Sep,Azreg-Ainou2014Mar,Azreg-Ainou2014May,Jusufi2019Aug}:
\begin{eqnarray}\label{trans_eq}
	dt&=&du+\frac{dr}{\sqrt{F(r)G(r)}},
\end{eqnarray}
where the functions $F(r)$ and $G(r)$ arises from the metric:
\begin{equation}
	ds^2 = -F(r)dt^2 + \frac{dr^2}{G(r)} + H(r)(d\theta^2 + \sin^2 \theta d\phi^2).
	\label{metric1}
\end{equation}
As per the notations in Ref. \cite{Jusufi2019Aug}, in our framework, we have $f(r) = F(r) = G(r)$ and $H(r) = r^2$.
The metric is conveniently reformulated using null tetrads as
\begin{equation}
	g^{\mu\nu} = -l^{\mu} n^{\nu} - l^{\nu} n^{\mu} + m^{\mu}\overline{m}^{\nu} + m^{\nu}\overline{m}^{\mu},
\end{equation}
where the null tetrads are defined by \cite{Jusufi2019Aug}
\begin{equation}
	\begin{aligned}
		l^{\mu} &= \delta^{\mu}_{r},\\ 
		n^{\mu} &= \delta^{\mu}_{u} - \frac{1}{2}F(r)\,\delta^{\mu}_{r},\\ 
		m^{\mu} &= \frac{1}{\sqrt{2H}}\left(\delta^{\mu}_{\theta} + \frac{\dot{\iota}}{\sin\theta}\,\delta^{\mu}_{\phi}\right),\\ 
		\overline{m}^{\mu} &= \frac{1}{\sqrt{2H}}\left(\delta^{\mu}_{\theta} - \frac{\dot{\iota}}{\sin\theta}\,\delta^{\mu}_{\phi}\right).
	\end{aligned}
\end{equation}

Notably, the tetrads are chosen so that $m^\mu$ and $\overline{m}^\mu$ are complex null vectors, and $\overline{m}^\mu$ represents the complex conjugate of $m^\mu$. In the Newman-Penrose formalism, this transformation to complex form is important because it makes it easier to break down spacetime geometry into a null tetrad basis that can better represent the directional properties of the gravitational field.
By definition, the set $\{l^\mu, n^\mu, m^\mu, \overline{m}^\mu\}$ constitutes a null tetrad that satisfies the orthogonality, normalization and isotropy conditions:
\begin{equation}
	\begin{aligned}
		l^{\mu}l_{\mu} = n^{\mu}n_{\mu} = m^{\mu}m_{\mu} = \overline{m}^{\mu}\overline{m}_{\mu} &= 0,\\ 
		l^{\mu}m_{\mu} = l^{\mu}\overline{m}_{\mu} = n^{\mu}m_{\mu} = n^{\mu}\overline{m}_{\mu} &= 0,\\ 
		-l^{\mu}n_{\mu} = m^{\mu}\overline{m}_{\mu} &= 1.
	\end{aligned}
\end{equation}
Except for the inner products $l^\mu n_\mu = -1$ and $m^\mu \overline{m}_\mu = 1$, which act as normalization conditions, these relations guarantee that all tetrad vectors are null and mutually orthogonal. 

Following the NJA, we can express the coordinate transformation as \cite{Azreg-Ainou2014Sep,Azreg-Ainou2014Mar,Azreg-Ainou2014May,Jusufi2019Aug} 
\begin{equation}
	{x'}^{\mu} = x^{\mu} + ia (\delta_r^{\mu} - \delta_u^{\mu}) \cos\theta,
	\label{NJA1}
\end{equation}
which yields the following relations:
\begin{equation}
\begin{cases}
	u' = u - ia\cos\theta, \\
	r' = r + ia\cos\theta, \\
	\theta' = \theta, \\
	\phi' = \phi,
\end{cases}
\end{equation}
where $a$ represents the spin parameter. Under this transformation, the null tetrad vectors $Z^\alpha$ transform according to
$
Z^\mu = \frac{\partial x^\mu}{\partial {x^\prime}^\nu}\, {Z^\prime}^\nu,
$
that results into \cite{Azreg-Ainou2014Sep,Azreg-Ainou2014Mar,Azreg-Ainou2014May,Jusufi2019Aug}:
\begin{equation}
	\begin{aligned}
		l'^{\mu} &= \delta^{\mu}_{r},\\ 
		n'^{\mu} &= \sqrt{\frac{B}{A}}\delta^{\mu}_{u} - \frac{1}{2}B\delta^{\mu}_{r},\\ 
		m'^{\mu} &= \frac{1}{\sqrt{2\,\Sigma}}\left[(\delta^{\mu}_{u}-\delta^{\mu}_{r})\dot{\iota}\,a\sin\theta + \delta^{\mu}_{\theta} + \frac{\dot{\iota}}{\sin\theta}\,\delta^{\mu}_{\phi}\right],\\ 
		\overline{m}'^{\mu} &= \frac{1}{\sqrt{2\,\Sigma}}\left[(\delta^{\mu}_{u}-\delta^{\mu}_{r})\dot{\iota}\,a\sin\theta + \delta^{\mu}_{\theta} + \frac{\dot{\iota}}{\sin\theta}\,\delta^{\mu}_{\phi}\right].
	\end{aligned}
	\label{nj_eqs}
\end{equation}
Here it is assumed that the functions $(G(r), F(r), H(r))$ transform to $(A(a,r,\theta), B(a,r,\theta), \Sigma(a,r,\theta))$. This notion was first proposed by Azreg-A\"{i}nou in Ref. \cite{Azreg-Ainou2014Sep}, where he developed a non-complexification procedure to generate rotating BH and non-BH solutions in GR. He showed that this technique is applicable to study the properties of Ay\'on--Beato--Garc\'ia BH. The technique is further extended to study wormholes with imperfect fluids with and without electric and magnetic fields \cite{Azreg-Ainou2014May} and in generating some general solutions of an imperfect fluid and its conformal homologous counterpart \cite{Azreg-Ainou2014Mar}. In Ref. \cite{Azreg-Ainou2014Mar}, the author stressed that the use of conformal fluids as cores of both static and rotating fluids does not suffer from malicious behaviour in terms of regularity of the solutions.

With the null tetrad vectors defined, the contravariant components of the new metric can be constructed as:
\begin{eqnarray}
	\begin{aligned}
		g^{uu} &= \frac{a^{2}\sin^{2}\theta}{\Sigma},  & g^{u\phi} &= \frac{a}{\Sigma},                  & g^{ur} &= 1-\frac{a^{2}\sin^2\theta}{\Sigma}, \\ 
		g^{rr} &= \mathcal{F}+\frac{a^{2}\sin^{2}\theta}{\Sigma}, & g^{r\phi} &= -\frac{a}{\Sigma},  & g^{\theta\theta} &= \frac{1}{\Sigma}, \\ 
		g^{\phi\phi} &= \frac{1}{\Sigma\sin^2\theta}.
	\end{aligned}
	\label{metric_terms}
\end{eqnarray}

Here, $\Sigma = r^2 + a^2 \cos^2\theta$ and $\mathcal{F}$ is a function of $r$ and $\theta$. The metric is then given by
\begin{align}
	ds^2 ={} & -\mathcal{F}\,du^2 - 2\,du\,dr + 2a\sin^2\theta\left(\mathcal{F}-1\right)du\,d\phi \nonumber\\ 
	&+ 2a\sin^2\theta\,dr\,d\phi + \Sigma\,d\theta^2 \nonumber\\ 
	&+ \sin^2\theta\left[\Sigma + a^2\left(2-\mathcal{F}\right)\sin^2\theta\right]d\phi^2.
\end{align}
Furthermore, the BH solution can be re-expressed in the original coordinates by applying the transformations
\begin{equation}
	du = dt - \frac{a^2+r^2}{\Delta}\,dr,\quad d\phi = d\varphi - \frac{a}{\Delta}\,dr,
\end{equation}
with $\Delta$ defined as \cite{Azreg-Ainou2014Sep,Azreg-Ainou2014Mar,Azreg-Ainou2014May,Jusufi2019Aug}
\begin{equation}
	\begin{aligned}
		\Delta &= r^{2}f(r) + a^{2} \\&= a^2-2 M r-\frac{4 \pi  r^2 \left(r_s+2 r\right) r_s^3 \rho _s}{3 \left(r_s+r\right){}^2}+r^2,
	\end{aligned}
	\label{delta_metric}
\end{equation}
where $f(r)$ is the lapse function of the NED BH assumed in our work. Notice that as, $\zeta = Q = 0$, the usual Kerr solution $\Delta = a^2 - 2 M r + r^2 $  is obtained.

\subsection{Shape of Ergoregion}

{
We now examine the structure of the ergoregion for a rotating BH located in a Dehnen-type galactic dark matter halo, as defined by the metric function in Eq.~\eqref{delta_metric}. The ergoregion is defined by the event horizon and the outer stationary limit surface, known as the ergosurface. It is a key feature of rotating BHs. Inside this region, no static observer can stay at rest because of the strong frame-dragging of spacetime. This effect is important in astrophysical processes like energy extraction and the formation of relativistic jets.

The horizons of the BH are located by the equation $\Delta = 0$:
\begin{equation}
a^2 - 2 M r + r^2 - \frac{4 \pi r^2 (r_s + 2 r) r_s^3 \rho_s}{3 (r_s + r)^2} = 0,
\end{equation}
where $a$ denotes the dimensionless spin parameter, $r_s$ the dark matter halo radius, and $\rho_s$ the central density of the Dehnen profile. Due to the complexity introduced by the dark matter terms, analytical solutions for the horizon radii are cumbersome, and we therefore rely on numerical methods for specific parameter choices.
\begin{figure*}[htb]
    \centering
    \includegraphics[scale = 0.50]{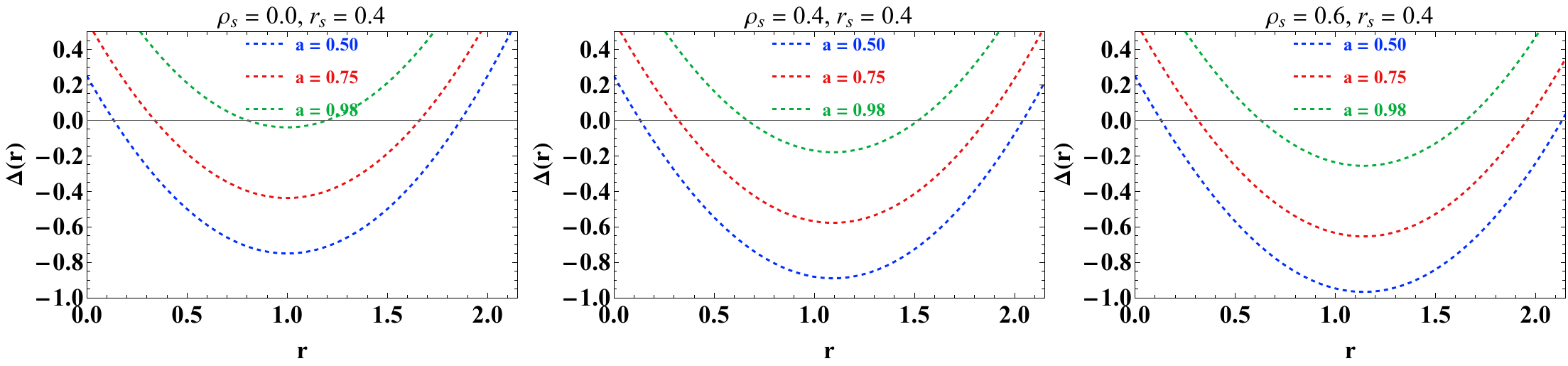}
    \includegraphics[scale = 0.50]{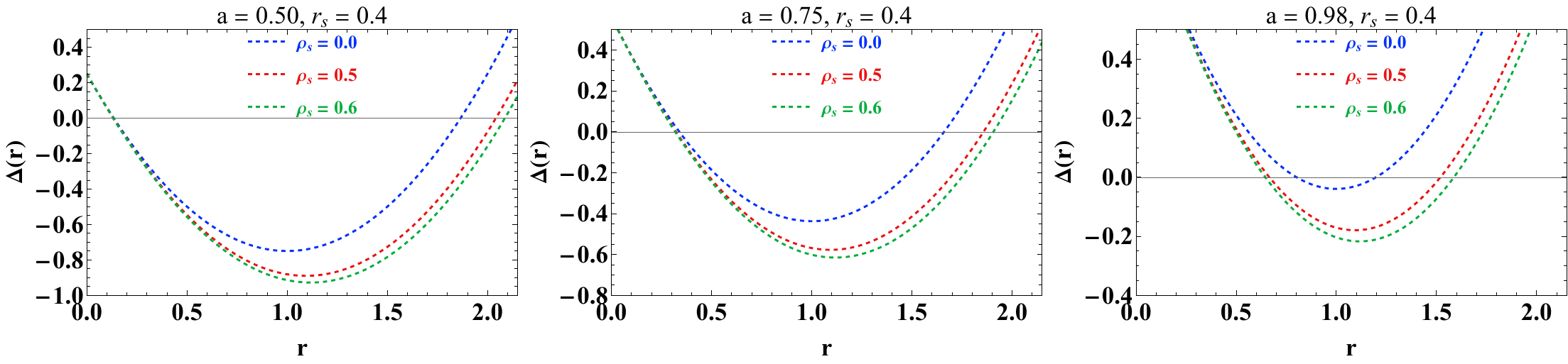}
    \includegraphics[scale = 0.50]{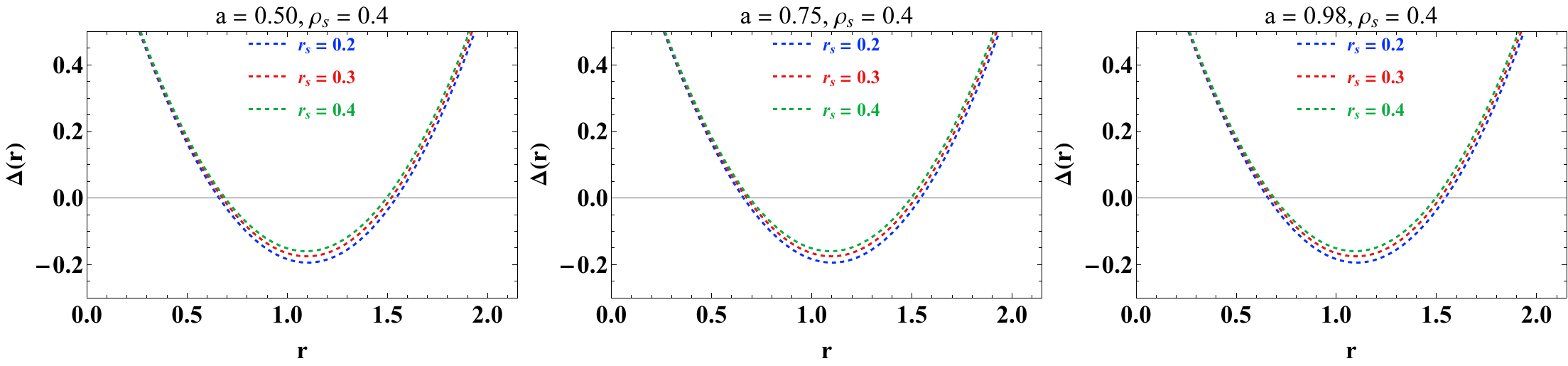}
    \caption{The plot of $\Delta(r)$ vs $r$ is shown for different choices of the spin parameters.}
    \label{delta_plot}
\end{figure*}	

Fig.~\ref{delta_plot} shows how the function $\Delta(r)$ behaves for different combinations of $a$, $r_s$, and $\rho_s$. The zeros of $\Delta(r)$ represent the BH horizons. Thus, these plots provide important insights into how spin and the parameters of the dark matter halo affect the horizon structure.

In the topmost panel of Fig.~\ref{delta_plot}, we keep the halo core radius fixed at $r_s=0.4$ and look at how $\Delta(r)$ changes with three different central densities: $\rho_s = 0.0, 0.4, 0.6$. We use spin values of $a = 0.50, 0.75, 0.98$. When $\rho_s=0$, the plots revert to the standard Kerr spacetime, showing the typical double-horizon structure. As $\rho_s$ increases, the entire $\Delta(r)$ curve shifts downwards, pushing the roots to larger radii. This means that an increase in dark matter density causes both the event horizon and the inner (Cauchy) horizon to move outward, resulting in a larger BH horizon size at fixed spin.
The central panels show the effects of varying $\rho_s$ while keeping $a$ and $r_s$ constant. Increasing $\rho_s$ from 0 to 0.6 shifts the horizons outward and deepens the minimum of $\Delta(r)$. Notably, at high values of $\rho_s$, the inner and outer horizons come closer together and may even merge. One can also see that the inner horizon does not change its position significantly. This suggests that the BH can approach an extremal limit or potentially lose its horizons altogether. This situation might relate to naked singularities in extreme environments dominated by dense dark matter halos.
The bottom panels illustrate the effect of increasing the halo radius $r_s$ (from 0.2 to 0.4) while keeping $a$ and $\rho_s$ constant. Similar to increasing $\rho_s$, larger core radius values shift the horizons outward, however the effect is not very pronounced. This reflects a larger region of gravitational influence due to the more extended dark matter distribution.
These results show a careful balance between BH spin and the characteristics of the nearby galactic dark matter halo in shaping horizon properties. While higher spin tends to lessen the radius gap between the horizons, moving either halo density or halo radius upward increases the horizon radii. These effects highlight the importance of the galactic environment when modeling strong gravity regions around astrophysical BHs.

The ergoregion boundaries - the ergosurfaces - are found by solving $g_{tt} = 0$:
\begin{equation}
a^2 \cos^2 \theta - 2 M r + r^2 - \frac{4 \pi r^2 (r_s + 2 r) r_s^3 \rho_s}{3 (r_s + r)^2} = 0.
\end{equation}
This condition, solved for each polar angle $\theta$, defines the angular dependence of the inner and outer ergosurfaces.
\begin{figure*}[htb]
\centerline{
\includegraphics[scale=0.3]{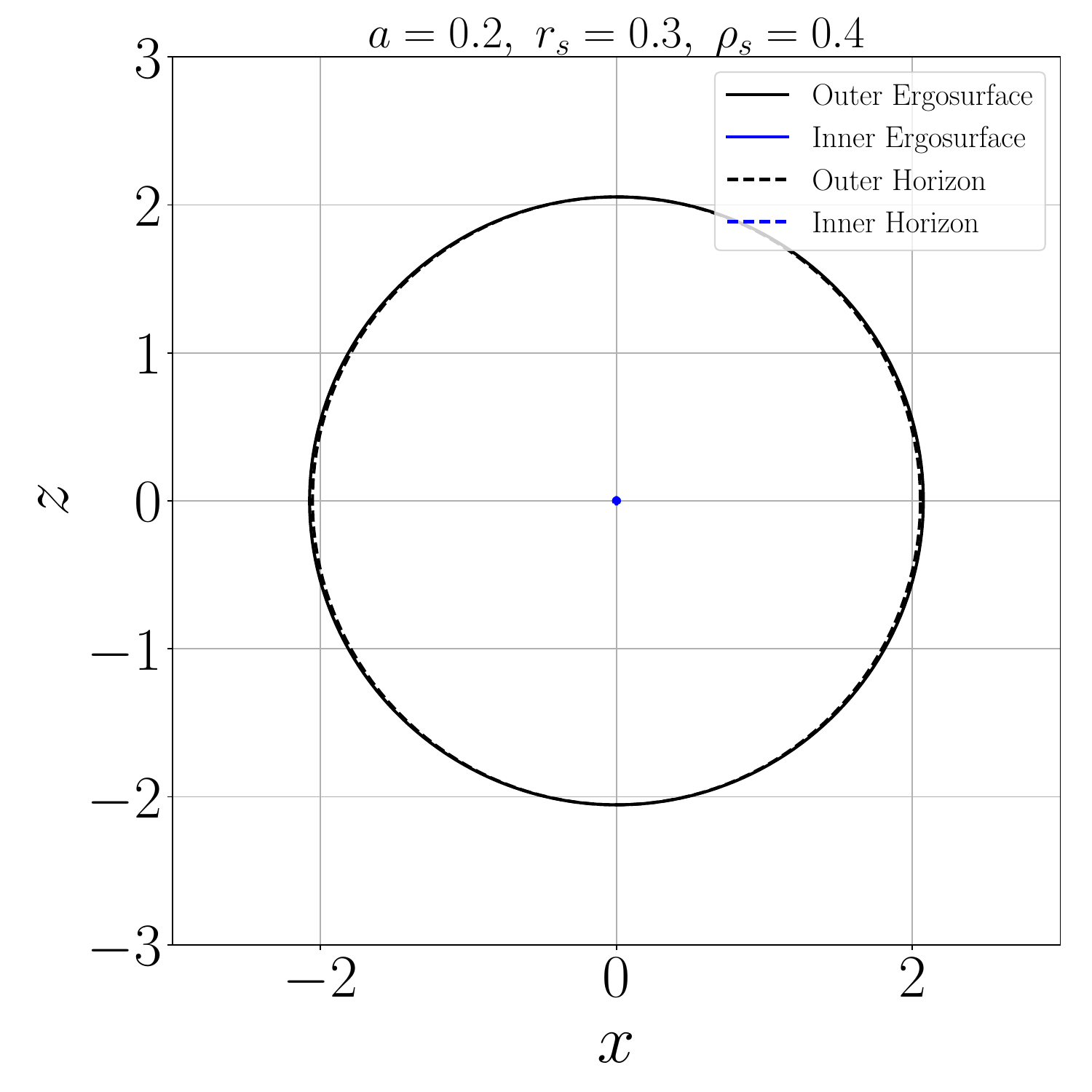}
\includegraphics[scale=0.3]{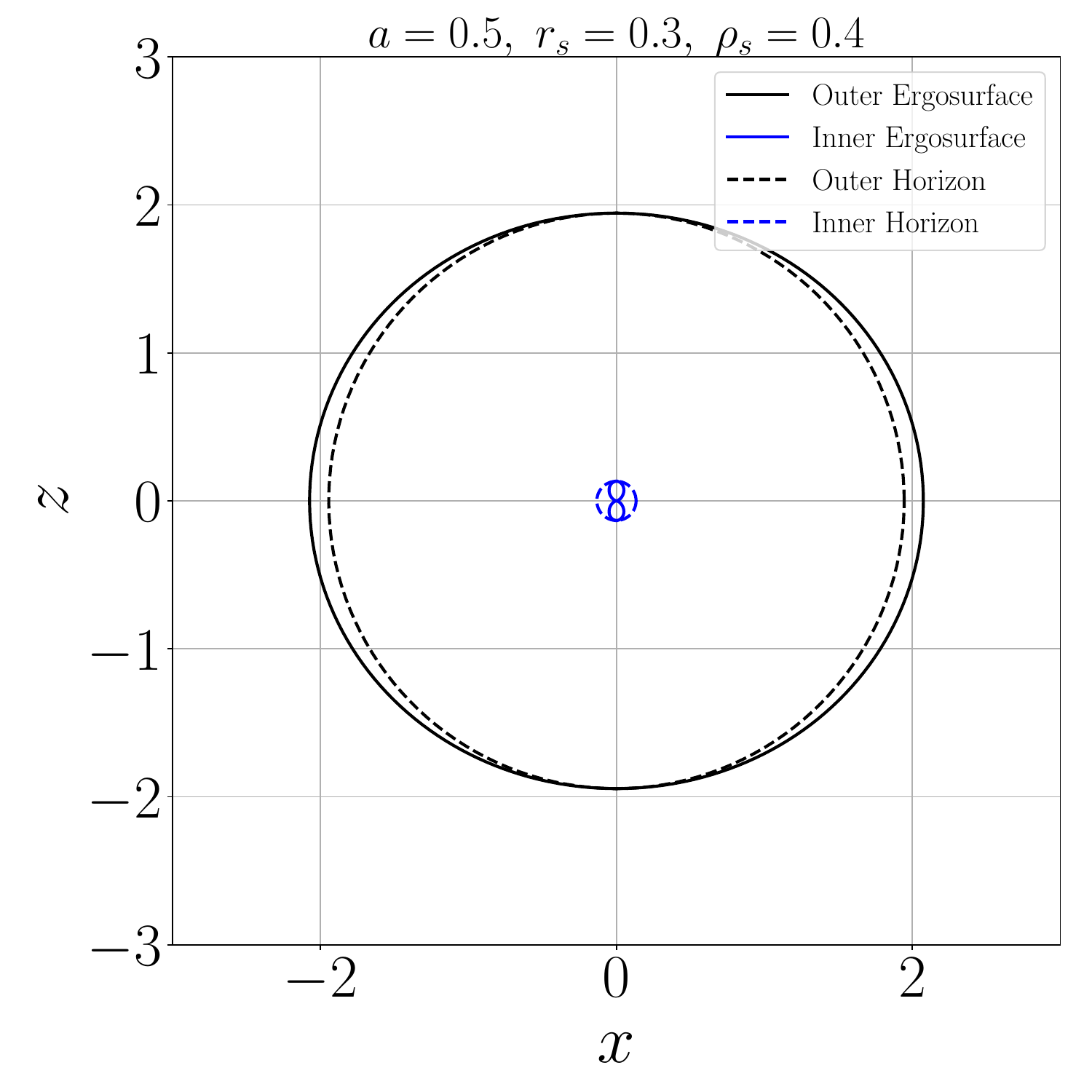}}
\centerline{
\includegraphics[scale=0.3]{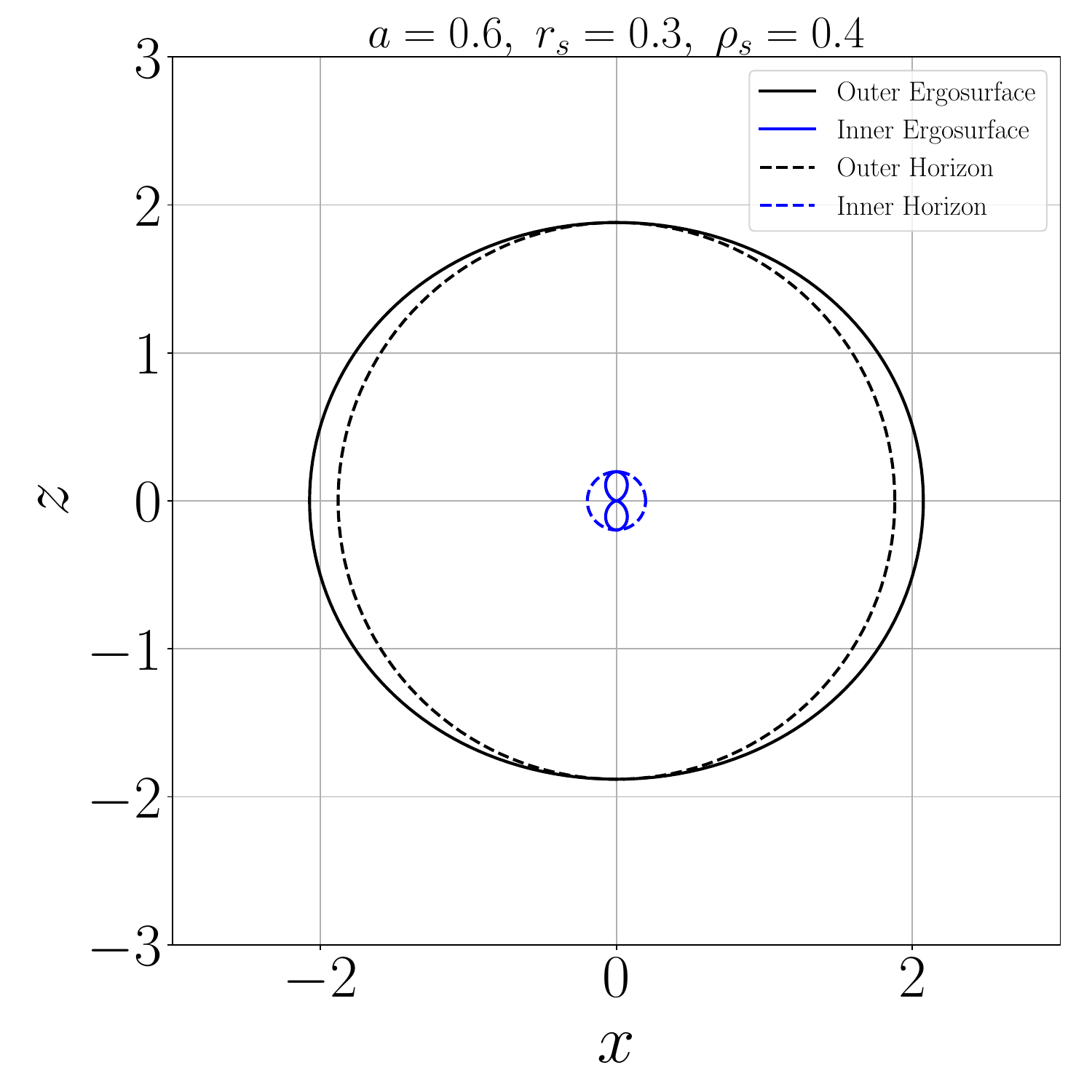}
\includegraphics[scale=0.3]{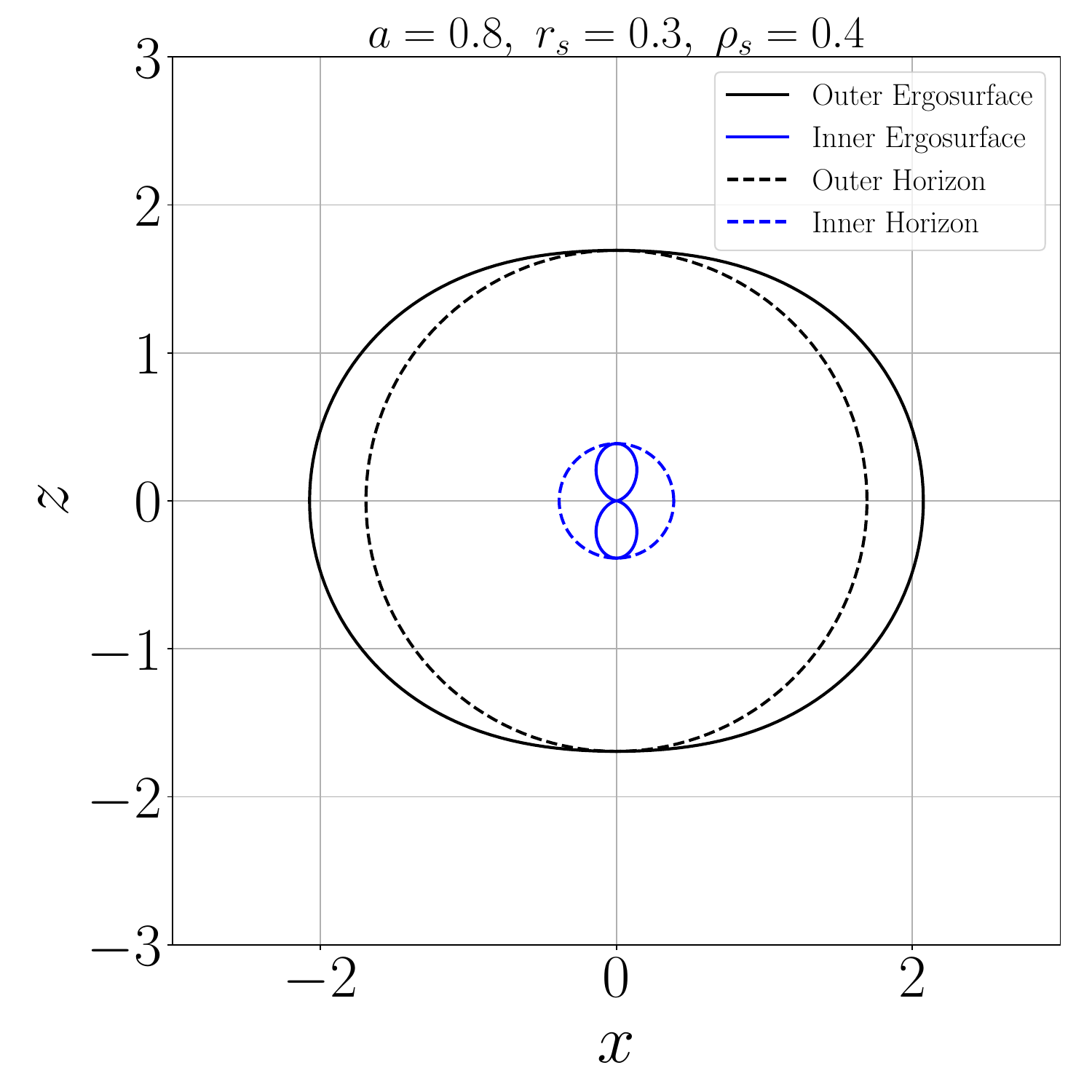}}
\centerline{
\includegraphics[scale=0.3]{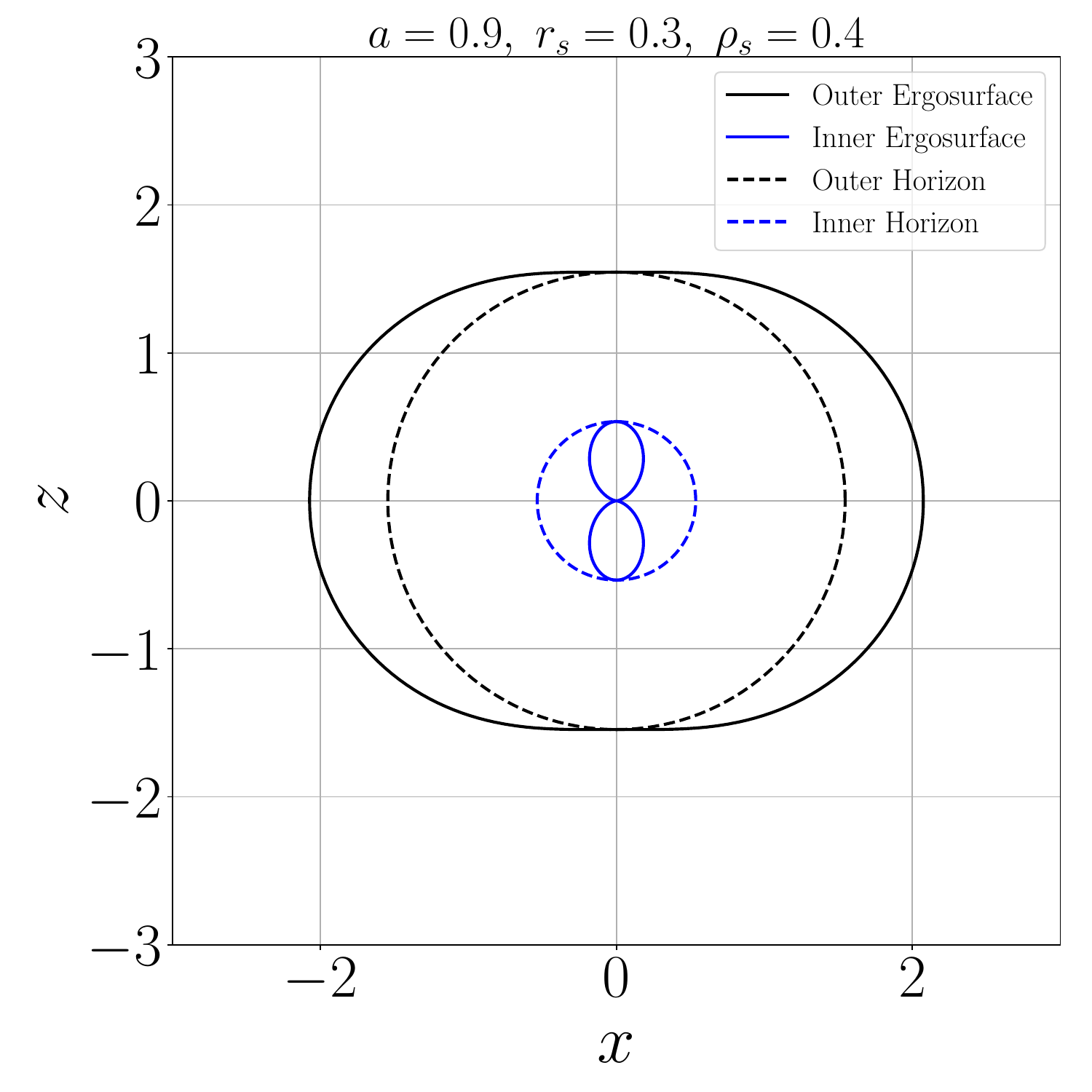}
\includegraphics[scale=0.3]{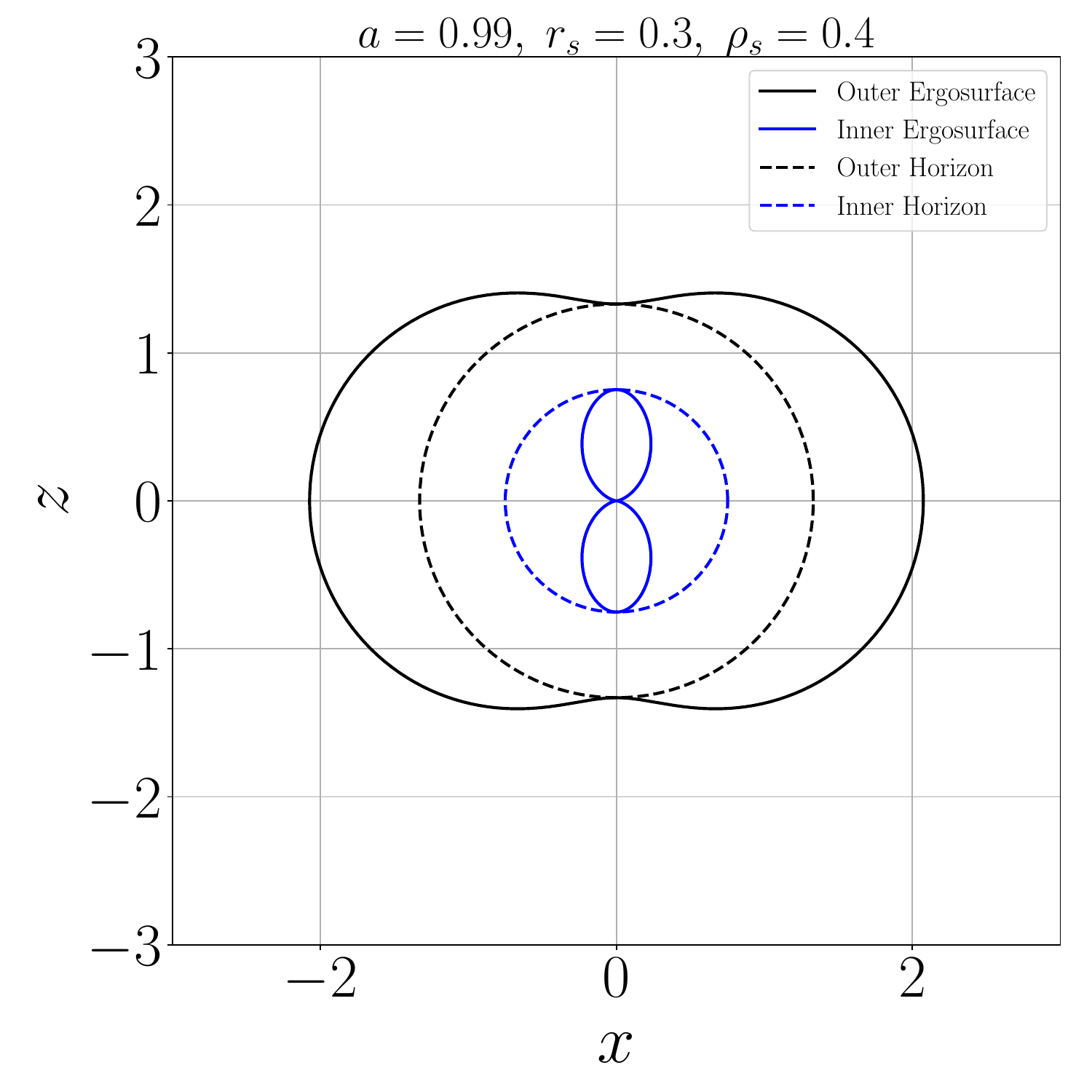}}
\caption{ Two-dimensional slices of the ergoregion in the $xz$-plane for various values of the spin parameter $a$, with fixed $r_s = 0.3$ and $\rho_s = 0.4$. Shown are the inner and outer ergosurfaces (static limits) as well as the inner and outer event horizons.}
\label{ergo_fig}
\end{figure*}

Fig.~\ref{ergo_fig} shows two-dimensional cross sections of the ergoregion in the $xz$-plane for different spins $a$, with the halo parameters set at $r_s=0.3$ and $\rho_s=0.4$. In each case, both the inner and outer horizons are displayed, along with their corresponding inner and outer ergosurfaces.
At low spin ($a=0.2$), the ergoregion is narrow. The outer ergosurface follows the event horizon closely, especially near the poles. As the spin increases, the ergoregion expands and becomes significantly distorted, particularly along the equatorial plane. In this region, the ergoregion bulges outward, increasing the space where rotational frame-dragging effects are strong.
For high spin values ($a=0.9$ and $a=0.99$), the outer ergosurface stretches significantly along the equator while remaining relatively spherical near the poles. This change in shape demonstrates the strong frame-dragging caused by rapid rotation, which increases the space available for energy extraction through the Penrose process and related astrophysical events.
Notably, the presence of the Dehnen dark matter halo increases both the horizons and the ergoregion compared to a typical Kerr BH. The additional gravitational pull from the halo pushes these surfaces outward, expanding the ergoregion and possibly increasing the potential for rotational energy extraction in galaxies surrounded by dark matter halos.
These plots clearly indicate that both BH spin and halo properties play a crucial role in shaping the geometry of the ergoregion. These findings are important for understanding jet formation, accretion dynamics, and high-energy activities near supermassive BHs in realistic galactic environments.

}

	\section{The Hamilton Jacobi Equation: Shadows}
	\label{sec4}
In this section, let us determine the shadow cast by the rotating NED BH. To do that, we first construct the geodesic equations of motion of photons. To obtain the null geodesics we follow the Hamilton-Jacobi formalism as follows: 
	\begin{equation}
		\partial_\tau\mathcal{J}=-\mathcal{H}.
		\label{HJ1}
	\end{equation}
	Here, $\mathcal{J}$ denotes the Jacobi action, which can be defined in terms of the affine parameter $\tau$ and the coordinates $x^\mu$ as $\mathcal{J}=\mathcal{J}(\tau,x^\mu)$ and $\mathcal{H}$ denotes the Hamiltonian of the particle, as given by $g^{\mu\nu}\partial_\mu\mathcal{J}\;\partial_\nu\mathcal{J}$. The spacetime symmetries dictates that photon energy $E$ and angular momentum $L$ must be conserved quantities, which are given by the usual definition of Killing fields $\kappa_t=\partial_t$ and $\kappa_\phi=\partial_\phi$, respectively. It is well known that the Hamilton-Jacobi equation can be solved in terms of separable solutions, which contains the already existing conserved quantities, i.e,
	\begin{equation}
		\mathcal{J}=\frac{1}{2}m^2\tau-Et+L\phi+\mathcal{J}_r(r)+\mathcal{J}_{\theta}(\theta)
		\label{HJ2}
	\end{equation}
	with $\mathcal{J}_r(r)$ and $\mathcal{J}_\theta(\theta)$ are functions of the coordinates $r$ and $\theta$.

	Combining Eq. (\ref{HJ1}) and Eq. (\ref{HJ2}) one can obtain the geodesic equations of motion expressed as the four velocity components as follows:
		\begin{align}
	\label{HJ3}
	&\Sigma\frac{dt}{d\tau}=\frac{r^2+a^2}{\Delta}[E(r^2+a^2)-aL]-a(aE\sin^2\theta-L),\\
	&\Sigma\frac{dr}{d\tau}=\sqrt{\mathcal{R}(r)},\\
	&\Sigma\frac{d\theta}{d\tau}=\sqrt{\Theta(\theta)},\\
	\label{HJ4}
	&\Sigma\frac{d\varphi}{d\tau}=\frac{a}{\Delta}[E(r^2+a^2)-aL]-\left(aE-\frac{L}{\sin^2\theta}\right),
	\end{align}
	where $\mathcal{R}(r)$ and $\Theta(\theta)$ are expressed by
	\begin{align}
	\label{HJ5}
	&\mathcal{R}(r)=[E(r^2+a^2)-aL]^2-\Delta[m^2r^2+(aE-L)^2+\mathcal{K}],\\
	&\Theta(\theta)=\mathcal{K}-\left(  \dfrac{L^2}{\sin^2\theta}-a^2E^2  \right) \cos^2\theta.
	\end{align}
	Here. $\mathcal{K}$ is a separation constant known as the Carter constant. This is an additional constant of motion arising from a hidden symmetry in rotating spacetimes, that ensures separability and full integrability of the geodesic equations of motion.

Depending on the value of the impact parameter, the photons from a light source may eventually be scattered away or fully captured by the BH. However, for some critical value of impact parameter, the photon may result in a photon sphere.  This behavior shows the area that delineates the edge of the shadow.  Using the effective potential, $V_{\text{eff}}$, related to the radial motion of the photon, to define the radial geodesic equation, we may investigate the existence of unstable circular orbits around the BH, which is defined by:
	\begin{equation}
	\Sigma^2\left(\frac{dr}{d\tau}\right)^2+V_{\text{eff}}=0.
	\end{equation}
	
Let us define two parameters $\xi$ and $\eta$ as \cite{Perlick2022Feb}:
	\begin{equation}
	\xi=L/E,  \quad \text{and} \quad   \eta=\mathcal{K}/E^2.
	\end{equation}
	
Also, these two parameters are directly related to $\Delta$ through \cite{Lambiase2025May}:
\begin{equation}
	\xi = \frac{\Delta_r'(r^2 + a^2) - 4\Delta_r r}{a \Delta_r'}, 
	\quad
	\eta = \frac{-r^4{\Delta_r'}^2 + 8r^3\Delta_r\Delta_r' + 16r^2\Delta_r(a^2 - \Delta_r)}{a^2 {\Delta_r'}^2}.
	\label{xi_eta_delta}
\end{equation}

	In terms of these two parameters, the effective potential can be redefined as:
	\begin{equation}\label{veff}
	V_{\text{eff}}=\Delta((a-\xi)^2+\eta)-(r^2+a^2-a\;\xi)^2.
	\end{equation}
	Here, we can safely replace $V_{\text{eff}}/E^2$ by $V_{\text{eff}}$ for convenience.\footnote{Keeping in mind that this scaling merely moves the peak of the effective potential to a higher or lower value without changing the position of the critical photon orbits.}. The critical photon orbits corresponding to a constant radius $r=r_c$ satisfies the following conditions:
	\begin{equation}\label{cond}
	V_{\text{eff}}(r)=0,\quad~~~\frac{dV_{\text{eff}}(r)}{dr}=0
	\end{equation}
	Using the conditions \eqref{cond}, we can obtain the parameters $\xi$ and $\eta$ for the rotating Schwarzschild-Dehnen BH as:
	\begin{widetext}
	\begin{equation}
	\xi(r)
= \frac{a^2 + r^2}{a}
- \frac{4\,r\bigl(a^2 - 2M r + r^2(1 - \alpha_3)\bigr)}
       {a\bigl(-2M + r^2(\alpha_1 - \alpha_2) + 2r(1 - \alpha_3)\bigr)}
	\label{xi_eq}
	\end{equation}
	\begin{equation}
\eta(r)
=\frac{%
  4\,M\,r^3\,(4a^2-9Mr)
  +r^5\bigl[\,8a^2(\alpha_1-\alpha_2)
    +(2(1-\alpha_3)-r(\alpha_1-\alpha_2))
     (12M+r^2(\alpha_1-\alpha_2)-2r(1-\alpha_3))
  \bigr]
}{
  a^2\bigl(-2M+r^2(\alpha_1-\alpha_2)+2r(1-\alpha_3)\bigr)^{2}
}
	\end{equation}
	\end{widetext}
 where 
\begin{align*}
\alpha_1(r) &= \frac{8\pi\,r_s^3\,(r_s+2r)\,\rho_s}{3\,(r_s+r)^3},\\
\alpha_2(r) &= \frac{8\pi\,r_s^3\,\rho_s}{3\,(r_s+r)^2},\\
\alpha_3(r) &= \frac{4\pi\,r_s^3\,(r_s+2r)\,\rho_s}{3\,(r_s+r)^2}.
\end{align*}

	In order to plot the shadow images, we need to use the coordinates $\alpha$ and $\beta$ which are defined through 
	
	\begin{equation}
		\begin{aligned}
		\alpha &= \lim_{r_0\to \infty} \left(-r_0^2 \sin \theta_0 \frac{d\phi}{dr} \Big|_{(r_0, \theta_0)}\right), \\ \beta &= \lim_{r_0\to \infty} \left(r_0^2 \frac{d\theta}{dr}\Big|_{(r_0, \theta_0)}\right).
		\end{aligned}
		\label{xy_shad}
	\end{equation}
	with $(r_0, \theta_0)$ denoting the observer's position.
	
	For any asymptotically flat spacetime, we take the $r \to \infty$ limit, which gives
	\begin{equation}
		\begin{aligned}
		\alpha &= - \frac{\xi}{\sin \theta_0}, \\
		\beta &= \pm \sqrt{\eta + a^2 \cos^2 \theta_0 - \xi^2 \cot^2 \theta_0}
		\end{aligned}
		\label{xy_shad2}
	\end{equation}
	
	When the observer is situated at the equatorial plane, one can set $\theta_0 = \pi/2$ which reduces the Eq. \eqref{xy_shad2} to 
	\begin{equation}
			\begin{aligned}
			\alpha &= - \xi, \\
			\beta &= \pm \sqrt{\eta}
		\end{aligned}
		\label{xy_shad_eq}
	\end{equation}
		
\begin{figure*}[htbp]
\centerline{\includegraphics[scale=0.30]{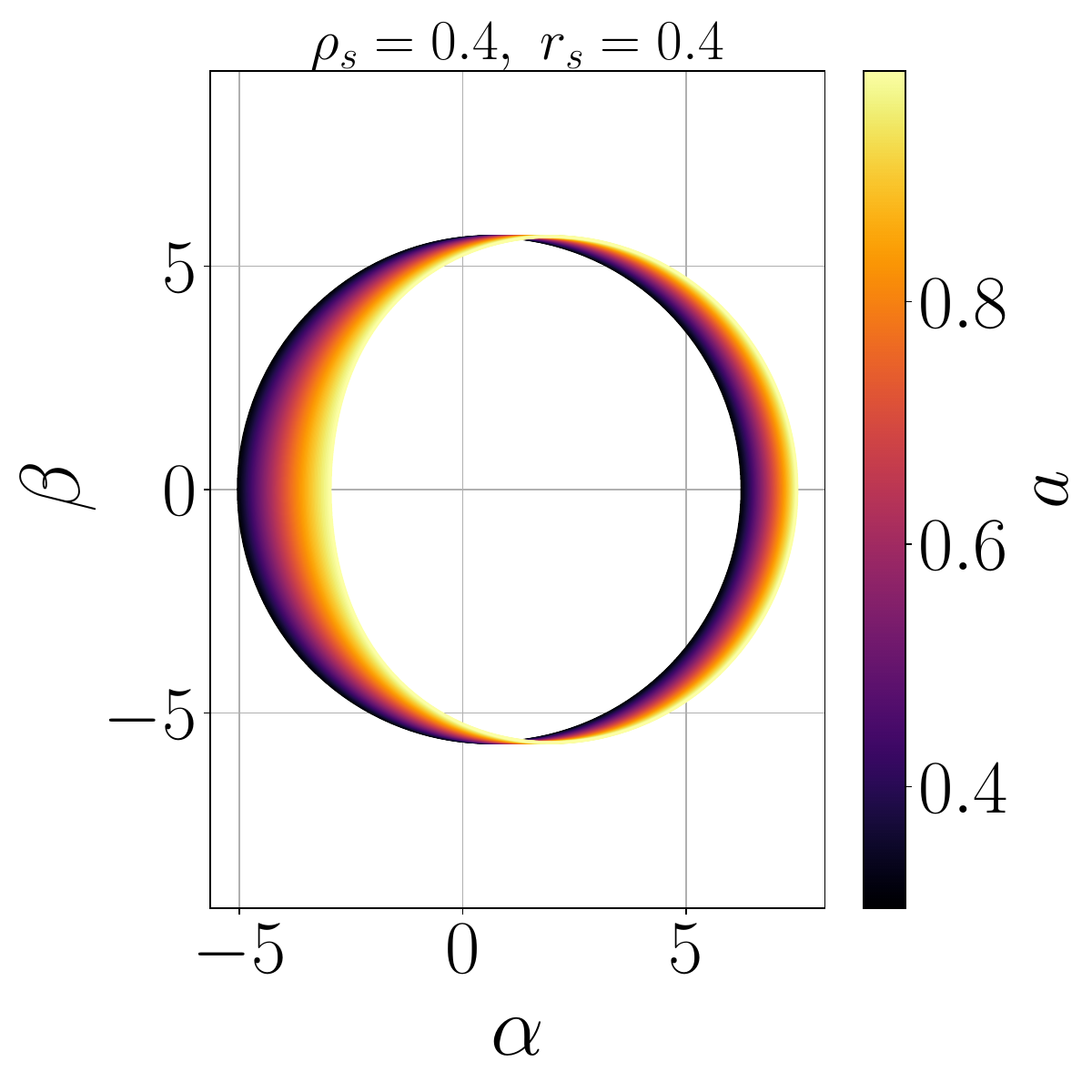}\includegraphics[scale=0.30]{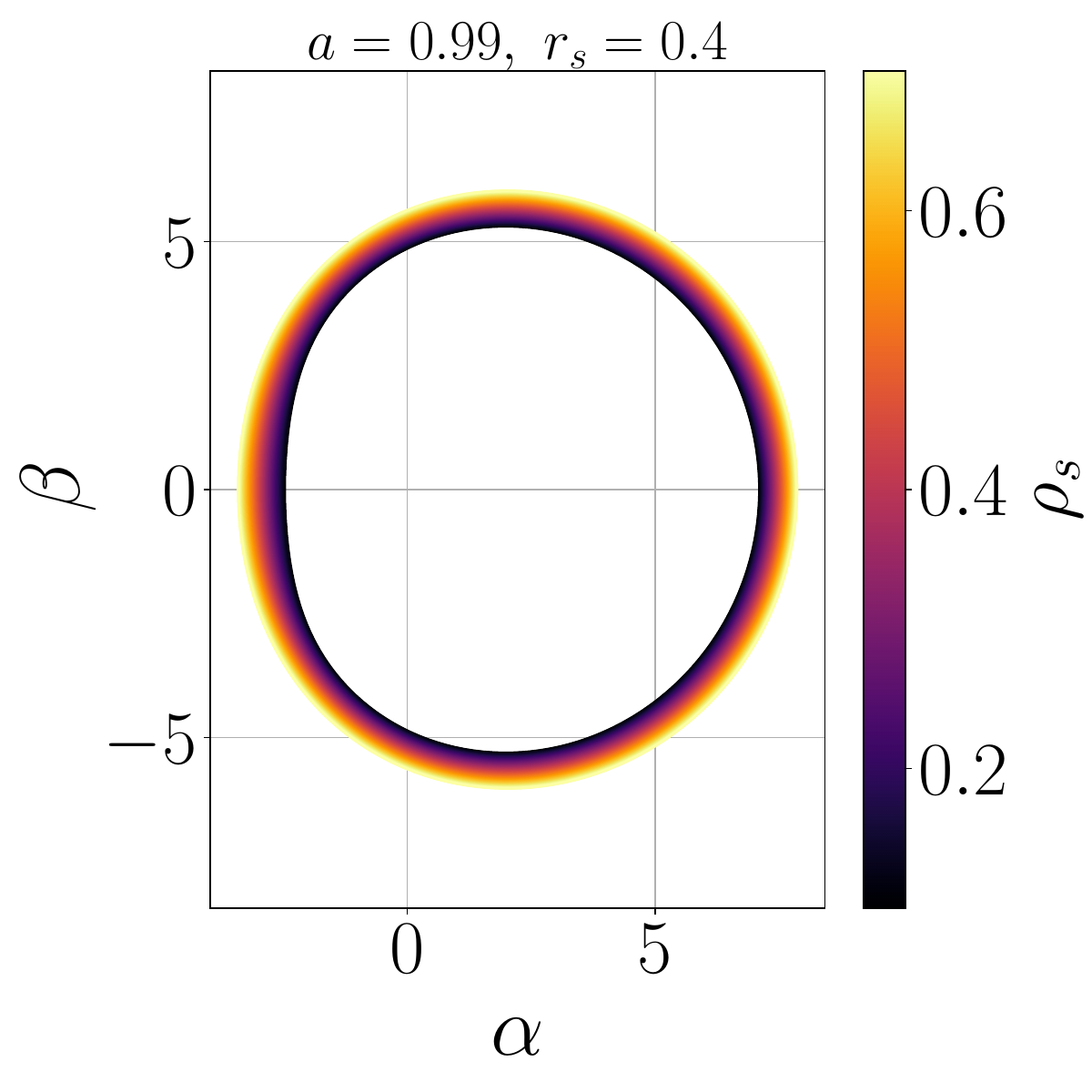}\includegraphics[scale=0.30]{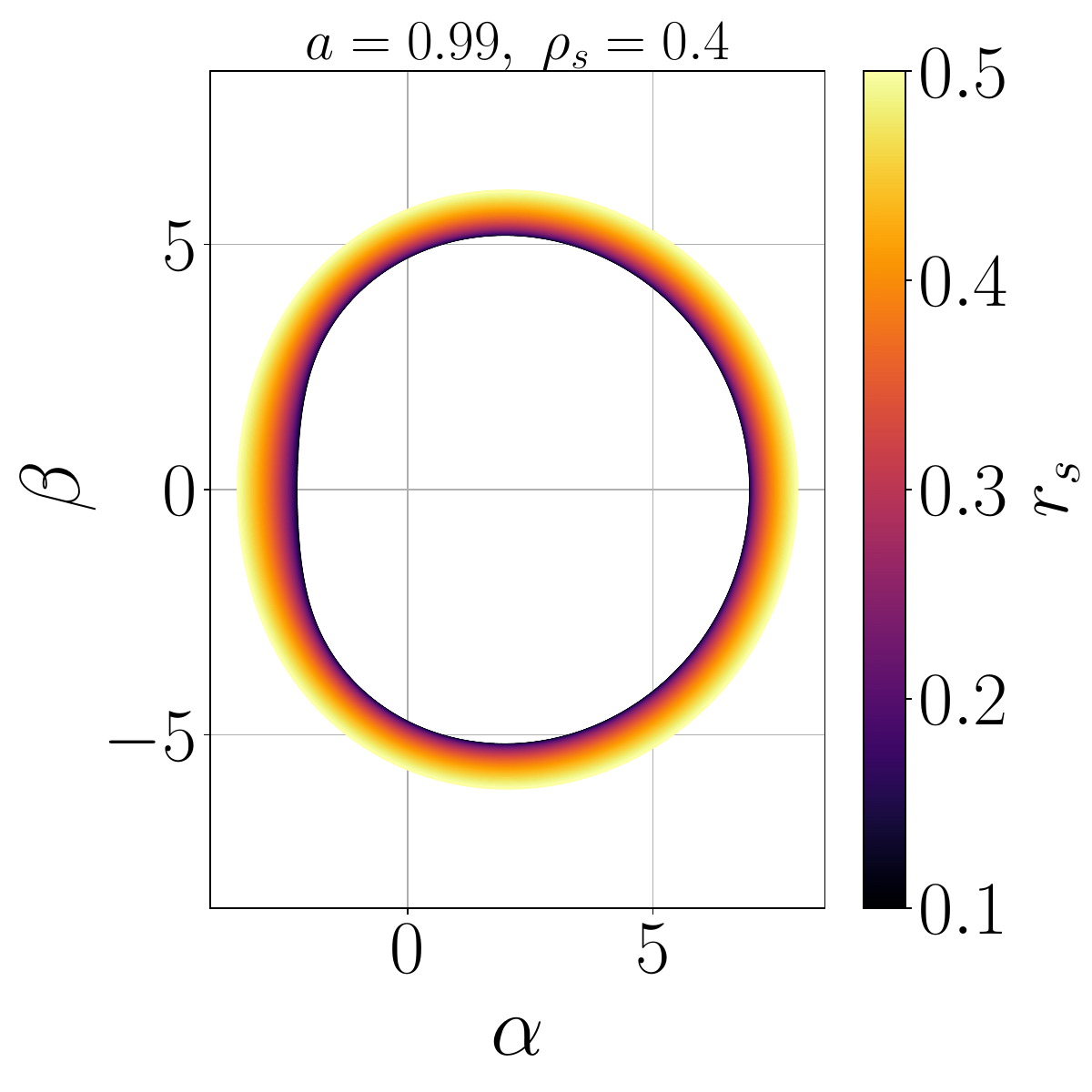}}
\caption{Shadows of the BH for different values of the model parameters.}
\label{shadowsfig}
\end{figure*}

{
Fig~\ref{shadowsfig} shows the BH shadow contours for different model parameters. It visually demonstrates how the properties of the dark matter halo and the BH spin influence observable shadow features. Each panel displays the shadow in the $(\alpha, \beta)$ plane, which corresponds to the celestial coordinates as viewed by a distant observer on the equatorial plane.
The left panel presents the shadow for a fixed spin of $a=0.4$ and a halo radius of $r_s=0.4$, while varying the central density $\rho_s$ of the Dehnen halo. As $\rho_s$ increases, the size of the shadow clearly grows. A higher dark matter core density increases the gravitational potential at larger radii. This strengthens the lensing effect and raises the effective photon sphere radius. Consequently, photons that come from farther away are more likely to be deflected into the BH, expanding the area of the sky where no signal can reach a distant observer. The thicker intensity ring, shown by the color bar, also indicates how dark matter density influences the photon ring structure.
The middle panel maintains the Dehnen halo parameters at fixed values ($r_s=0.4$, $\rho_s=0.4$) while varying the BH spin parameter $a$. As $a$ approaches its maximum value ($a=0.99$), the shadow becomes noticeably uneven, developing a flattened edge on the left. This distortion is a sign of frame dragging. In the prograde direction (left side), co-rotating photons can orbit closer to the event horizon because of the BH’s rotation. This reduces the shadow’s size in that direction, while the retrograde side expands outward. Such asymmetry results from the connection between photon movement and the BH's angular momentum. In real observations, noticing this asymmetry would provide strong evidence for BH spin and the underlying metric, as well as the presence of surrounding matter.
The right panel examines how varying the Dehnen halo radius $r_s$ impacts the shadow, while keeping $a=0.99$ and $\rho_s=0.4$ constant. As $r_s$ increases, the shadow once again expands. This shows that both the central density and the spatial extent of the halo play important roles in light bending and shadow formation. Larger halos change the spacetime geometry at greater distances, which shifts the photon sphere outward and increases the shadow’s diameter. The color gradient in all panels, which indicates the relative intensity of the photon ring, reflects the properties of photon trajectories near the critical curve. It may also be affected by both rotational dragging and halo gravitational lensing.
These results highlight two main points. First, the shadow size and distortion depend not just on the BH parameters, but also on the galactic dark matter environment, which makes significant changes. Second, future high-resolution observations, such as those by the Event Horizon Telescope and its successors, could potentially use shadow characteristics to infer BH properties. They could also help understand the nature and distribution of dark matter in galactic centers. This connection emphasizes the importance of considering environmental effects in modeling and analyzing BH images and sets the stage for multi-messenger constraints on dark matter and strong field gravity.
}

\subsection{Distortion}

The distortion parameter measures the deviation of a BH shadow from perfect circularity, thereby capturing the effects of rotation and other parameters on the shadow morphology. For static (non-rotating) BHs, the shadow is exactly circular, and the distortion vanishes. In contrast, rotating BHs produce an asymmetric shadow boundary, which requires a quantitative description of distortion.

The distortion is defined in terms of the linear radius of the shadow \cite{Hioki:2009na,Amir:2016cen,Raza:2023vkn}. This represents the radius of a hypothetical reference circle that passes through three characteristic points on the shadow: the topmost $(\alpha_t,\beta_t)$, the bottommost $(\alpha_b,\beta_b)$, and the rightmost $(\alpha_r,0)$. The linear radius is given by
\begin{equation}
R_{sh} = \frac{(\alpha_t - \alpha_r)^2 + \beta_t^2}{2|\alpha_t - \alpha_r|}. \label{eqlin_radius}
\end{equation}

Here, $(\alpha,\beta)$ denote the celestial coordinates of any point on the shadow contour, while the subscripts $t$, $b$, and $r$ correspond to the top, bottom, and right extremities, respectively. For a static BH, the shadow also passes through the leftmost point $(-\alpha_r,0)$, forming a perfect circle of radius $R_{sh}$.

For rotating BHs, the shadow is asymmetric, and the leftmost point on the boundary does not coincide with $(-\alpha_r,0)$. The degree of asymmetry is quantified by
\begin{equation}
\delta_s = \frac{|\bar{\alpha}_l - \alpha_l|}{R_{sh}}, \label{eqdistortion}
\end{equation}
where $(\alpha_l,0)$ represents the leftmost point of the actual shadow, and $(\bar{\alpha}_l,0)$ is the corresponding point on the reference circle. The subscript $l$ indicates the left side of the $\beta$-axis, and the bar refers to the reference circle.

Analysis of $\delta_s$ as a function of BH parameters provides key insights into the shadow deformation. Typically, $\delta_s$ increases with spin and reaches its maximum in the extremal limit, while additional charges or modified gravity terms can alter this trend depending on their contributions.

\begin{figure*}[htbp]
\centerline{\includegraphics[scale=0.730] {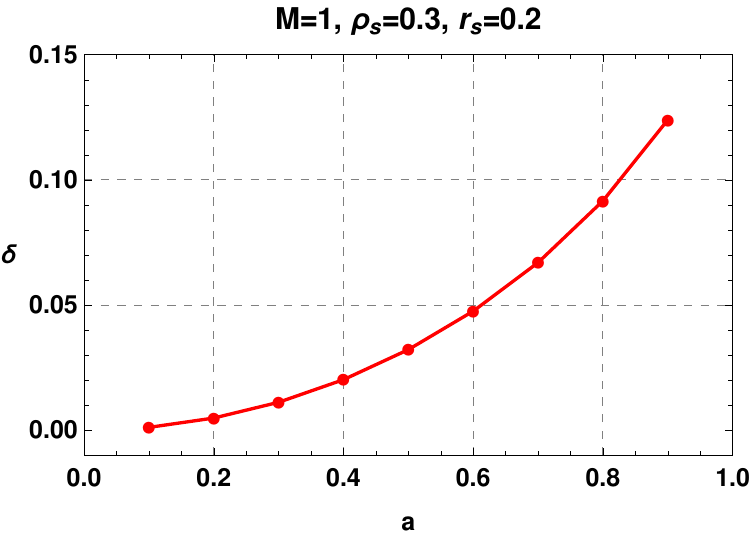}\hspace{2mm}\includegraphics[scale=0.760]{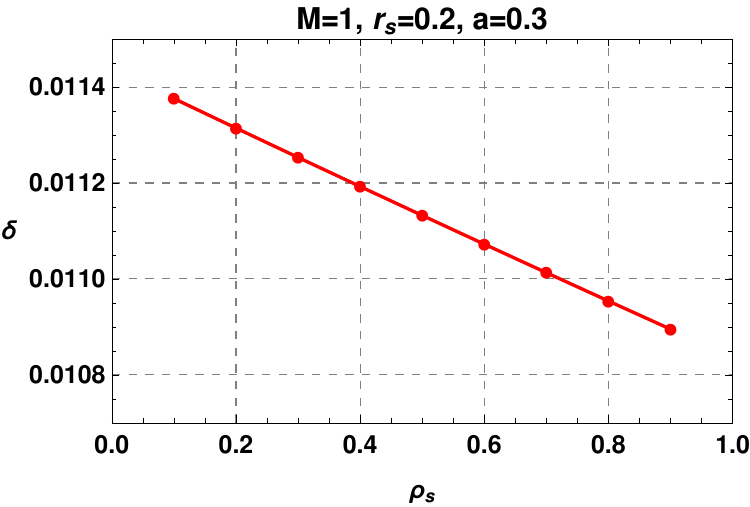}}\includegraphics[scale=0.730]{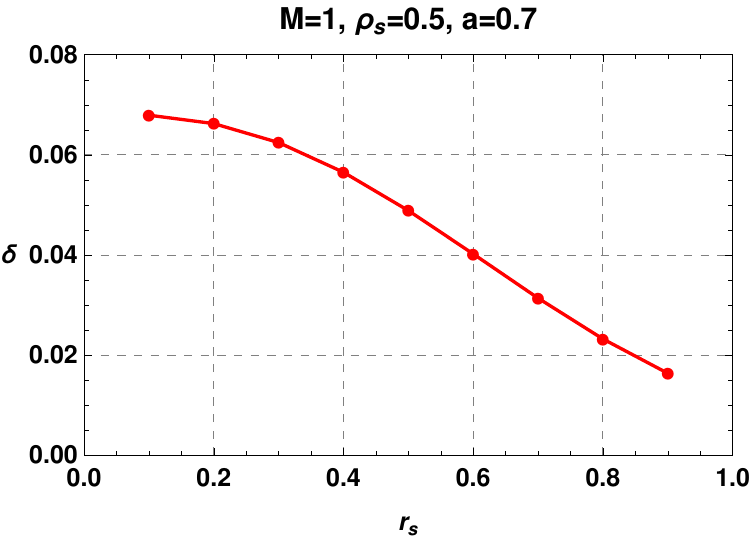}
\caption{Distortion of the BH for different values of the model parameters.}
\label{dist}
\end{figure*}

To understand how distortion is affected by the model parameters in our case, we have plotted distortion vs. model parameters in Fig. \ref{dist}. On the first panel of Fig. \ref{dist}, one can see that with an increase in the spin parameter $a$, the distortion of the BH shadow increases. On the second panel, one can see that $\rho_s$ has a linear impact on the distortion parameter $\delta$. With an increase in the value of $\rho_s$, distortion $\delta$ decreases linearly. However, the impact of $\rho_s$ on the distortion $\delta$ is very small. On the third panel, we have depicted the variation of $\delta$ w.r.t. $r_s$. With an increase in the value or $r_s$, $\delta$ decreases non-linearly.

\section{Energy Emission from Black Holes}\label{energyemission}
In classical physics, any object crossing a BH's event horizon is irretrievably lost. However, quantum mechanics suggests that BHs emit particles through a process tied to quantum fluctuations within the horizon. These fluctuations lead to particle-antiparticle pair creation, where particles with positive energy can escape via quantum tunneling, carrying away energy and contributing to the BH's evaporation. The absorption probability is quantified by the absorption cross-section. At large distances from the BH’s gravitational influence, this cross-section corresponds to the BH’s shadow. In the high-energy limit, the absorption cross-section stabilizes at a constant value, $\sigma_{lim}$, approximately equal to the area of the BH’s shadow, as given by \cite{Raza:2023vkn, Decanini:2011xi}:
\begin{equation} \sigma_{lim} \approx \pi R_{sh}^2, \end{equation}
where $R_{sh}$ is the radius of the shadow. The rate of energy emission is described by:
\begin{equation} \mathcal{E}{\omega t} := \frac{d^2\mathcal{E}(\omega)}{d\omega dt} = \frac{2\pi^2\sigma{lim}\omega^3}{e^{\omega/T_H} - 1} \approx \frac{2\pi^3 R_{sh}^2 \omega^3}{e^{\omega/T_H} - 1}, \end{equation}
where $\omega$ represents the angular frequency and $T_H = \kappa / 2\pi$ denotes the Hawking temperature. The surface gravity, $\kappa$, at the event horizon of a rotating BH is:
\begin{equation} \kappa = \frac{\Delta'(r)}{2(a^2 + r^2)}\bigg|_{r=r_h}, \end{equation}
where $r_h$ is the event horizon radius and $a$ is the BH’s spin parameter. For a non-rotating (static) BH, the surface gravity simplifies to:
\begin{equation} \kappa = \frac{1}{2} f'(r)\bigg|_{r=r_h}. \end{equation}

\begin{figure*}[htbp]
\centerline{\includegraphics[scale=0.830] {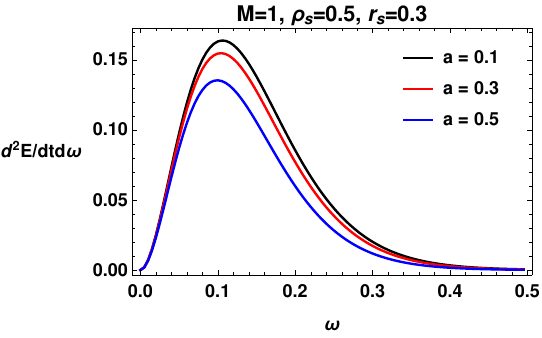}
\hspace{2mm}\includegraphics[scale=0.830]{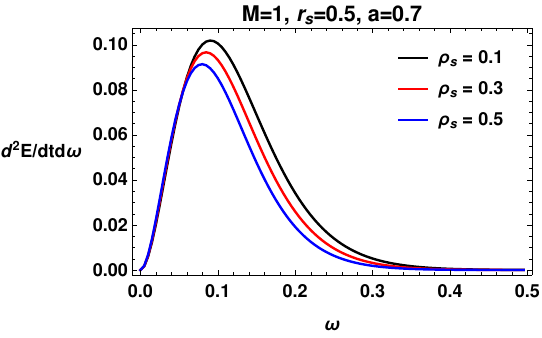}}\includegraphics[scale=0.830]{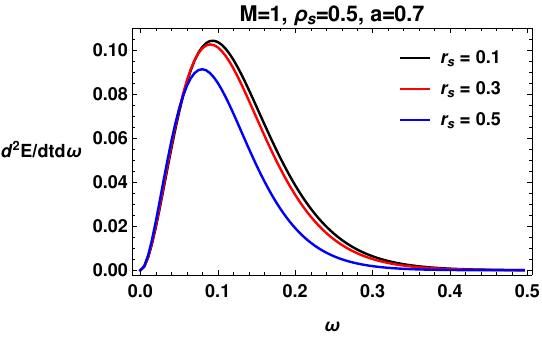}
\caption{Emission rate of the BH for different values of the model parameters.}
\label{emission}
\end{figure*}

In Fig. \ref{emission}, we have shown the variation of emission rate w.r.t. $\omega$ for different model parameters. One can see that with an increase in the value of model parameters $a$, $\rho_s$ and $r_s$, the emission rate of the BH decreases and as a result, the lifetime of the BH increases. It is worth to note that $a$ and $r_s$ have non-linear impacts on the emission rate of the BH.

\section{QNMs Using WKB Approximation} \label{sec5}

In this section, we present a comprehensive analysis of the QNMs of a rotating BH influenced by a Dehnen-type galactic dark matter halo. The study focuses exclusively on the rotating case, as it is most relevant for astrophysical BHs. We employ the Wentzel-Kramers-Brillouin (WKB) approximation technique \cite{Iyer:1986np, Dias:2022oqm, Konoplya:2019hlu, Konoplya:2011qq,Konoplya:2017wot,Konoplya:2003ii,Yang:2012he} to determine the complex frequencies characterizing the BH’s response to scalar perturbations. The Dehnen halo introduces additional gravitational influence through the core density $ \rho_s $ and halo radius $ r_s $, and we analyze how these parameters may affect the QNM spectra. The results offer critical insight into how the surrounding dark matter distribution modulates the dynamic behavior of rotating BHs and the possible observable imprints on gravitational wave signals.

QNMs describe damped oscillations of BHs under perturbations. These are characterized by complex frequencies $ \omega = \omega_R - i\omega_I $, where $ \omega_R $ represents the oscillation frequency and $ \omega_I $ indicates the damping rate. These modes are key signatures in gravitational wave astronomy, particularly in the ringdown phase following BH mergers. The detection and interpretation of such signals depend on a detailed understanding of how environmental factors, such as surrounding matter, alter these frequencies. Therefore, incorporating the influence of a galactic halo modeled by the Dehnen profile becomes essential.

Teukolsky showed that perturbations of various spins - scalar, vector, and tensor - in Kerr spacetime obey a unified master equation formulated for variables with spin weight $\bar{s}$. This equation admits separable solutions through an ansatz that decouples the radial and angular dependencies \cite{Teukolsky:1972my}.
The rotating BH metric in the presence of a Dehnen halo remains axially symmetric and stationary. The scalar field perturbation $ u(t, r, \theta, \phi) $ satisfies the Teukolsky master equation, which allows for a separable ansatz \cite{Luna:2022rql, Yang:2012he, Yang:2021zqy}:
\begin{equation} u(t, r, \theta, \phi) = e^{-i\omega t} e^{im_l \phi} u_r(r) u_\theta(\theta). \end{equation}
The angular part yields a generalized spheroidal harmonics equation:
\begin{equation} \frac{1}{\sin\theta} \frac{d}{d\theta} \left( \sin\theta \frac{du_\theta}{d\theta} \right) + \left[ a^2 \omega^2 \cos^2\theta - \frac{m_l^2}{\sin^2\theta} + \mathcal{A}_{lm_l} \right] u_\theta = 0, \end{equation}
where $ \mathcal{A}_{lm_l} $ is the angular eigenvalue. This equation, in the large multipole limit $ l \gg 1 $, is tackled using the WKB method. Applying the Bohr-Sommerfeld quantization yields \cite{Yang:2012he}:
\begin{equation} \int_{\theta_-}^{\theta_+} \sqrt{a^2 \omega^2 \cos^2\theta - \frac{m_l^2}{\sin^2\theta} + \mathcal{A}_{lm_l}} \, d\theta = \left( l + \frac{1}{2} - |m_l| \right) \pi. \end{equation}
Solving this in the eikonal limit, we find an approximate form:
\begin{equation} \mathcal{A}_{lm_l} \approx (l + 1/2)^2 - \frac{a^2 \omega^2}{2} \left( 1 - \frac{m_l^2}{(l + 1/2)^2} \right). \end{equation}
This relation demonstrates that the angular eigenvalue becomes complex when $ \omega $ is complex. Using perturbation theory, we derive the imaginary part:
\begin{equation} \mathcal{A}_{lm_l}^I = -2a^2\omega_R\omega_I \langle \cos^2\theta \rangle, \end{equation}
where the expectation value is computed using the WKB form of $ u_\theta $. This connects the damping rate to the angular behavior of the field, which is subtly influenced by the halo-modified frame dragging.

The radial Teukolsky equation, recast into a Schrödinger-like form using the tortoise coordinate $ r_* $, is given by:
\begin{equation} \frac{d^2 u_r}{dr_*^2} + V^r(r, \omega) u_r = 0, \end{equation}
with $ \frac{d}{dr_*} = \frac{\Delta}{r^2 + a^2} \frac{d}{dr} $. The effective potential $ V^r(r, \omega) $ includes contributions from the Dehnen halo through $ \Delta $. Explicitly,
\begin{equation} 
V^r(r, \omega) = \frac{\left(a m_l-\omega  \left(a^2+r^2\right)\right){}^2-\left(\frac{1}{2} a \omega  \left(4 m_l \left(\frac{a \omega  m_l}{(2 l+1)^2}-1\right)+a \omega \right)+l^2+l+\frac{1}{4}\right) \left(a^2-2 M r-\frac{4 \pi  r^2 \left(r_s+2 r\right) r_s^3 \rho _s}{3 \left(r_s+r\right){}^2}+r^2\right)}{\left(a^2+r^2\right)^2}. 
\end{equation}
To determine the QNM frequencies, we locate the maximum of $ V^r $ and apply the WKB quantization condition at the peak $ r_0 $. At leading order, the conditions are:
\begin{equation} 
V^r(r_0, \omega_R) = 0, \quad \left. \frac{\partial V^r}{\partial r} \right|_{r_0, \omega_R} = 0. 
\end{equation}
Solving these gives:
\begin{equation} 
\omega_R = \frac{a m_l \left(\left(r_s+r_0\right){}^3 \left(B_1+3 r_s-9 r_0\right)-4 \pi  r_s^3 \left(-6 r_0^2 r_s+r_s^3-2 r_0^3\right) \rho _s\right)}{\left(r_s+r_0\right){}^3 \left(B_1 \left(a^2-3 r_0^2\right)+3 \left(a^2-3 r_0^2\right) r_s+9 r_0 \left(a^2+r_0^2\right)\right)+4 \pi  B_2 r_s^3 \rho _s}, 
\end{equation}
where $B_1 = 9 M+4 \pi  r_s^3 \rho _s-3 r_s $ and $B_2 = 2 r_0^3 \left(a^2+3 r_s^2\right)+3 r_0^2 \left(2 a^2 r_s+r_s^3\right)-a^2 r_s^3-6 r_0^5. $
This expression highlights the dependence of the real frequency on the BH’s spin and the halo-induced geometry.

The damping rate is obtained from the imaginary part using:
\begin{equation} \omega_I = -\left(n + \frac{1}{2}\right) \frac{\sqrt{2 \left( \frac{d^2 V^r}{dr_*^2} \right)_{r_0, \omega_R}}}{\left(\frac{\partial V^r}{\partial \omega} \right)_{r_0, \omega_R}}. \end{equation}
This formulation gives the overtone-dependent decay rate. When $ \mathcal{A}_{lm_l} $ is substituted in terms of $ \omega $, and halo modifications are included, the expression becomes analytically cumbersome but numerically tractable.

\begin{figure}[htbp]
\centerline{
   \includegraphics[scale = 0.58]{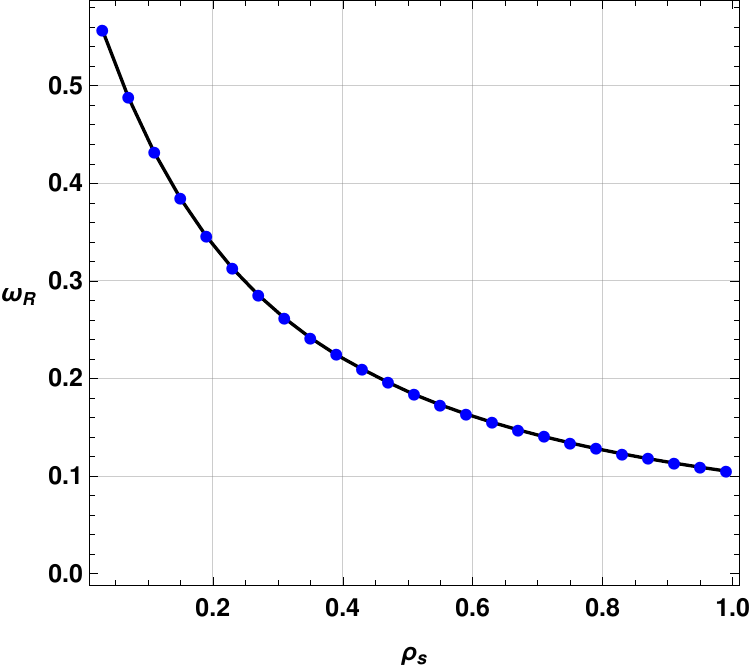}\hspace{0.5cm}
   \includegraphics[scale = 0.58]{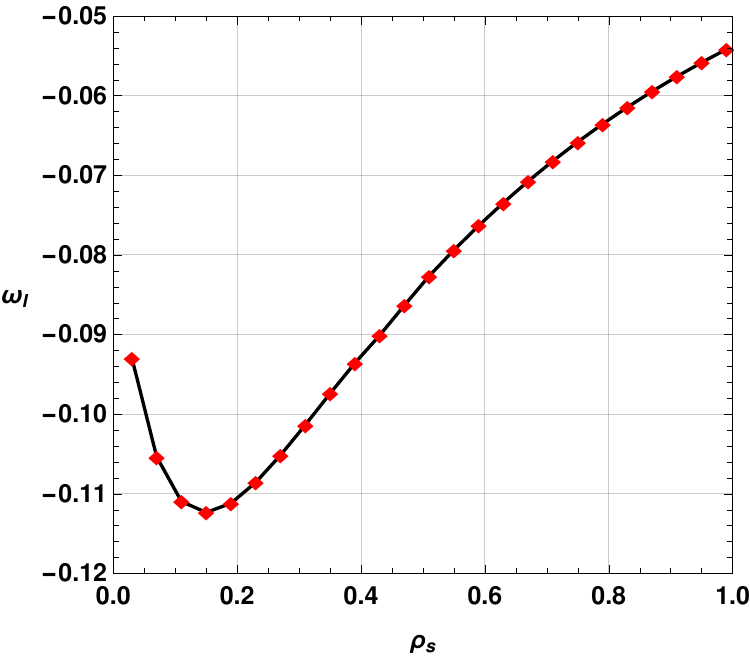}} \vspace{-0.2cm}
\caption{Variation of QNMs with respect to model parameter $\rho_s$
with $M=1$, $n= 0$, $r_s =1$, $a = 0.9$, $m_l=1$ and $l=2$ for scalar
perturbation.}
\label{QNMs01}
\end{figure}

\begin{figure}[htbp]
\centerline{
   \includegraphics[scale = 0.58]{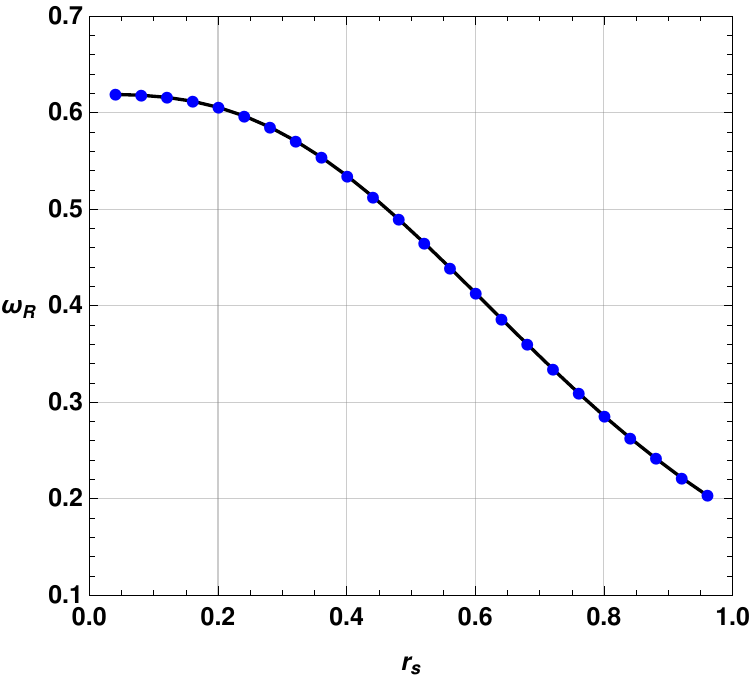}\hspace{0.5cm}
   \includegraphics[scale = 0.58]{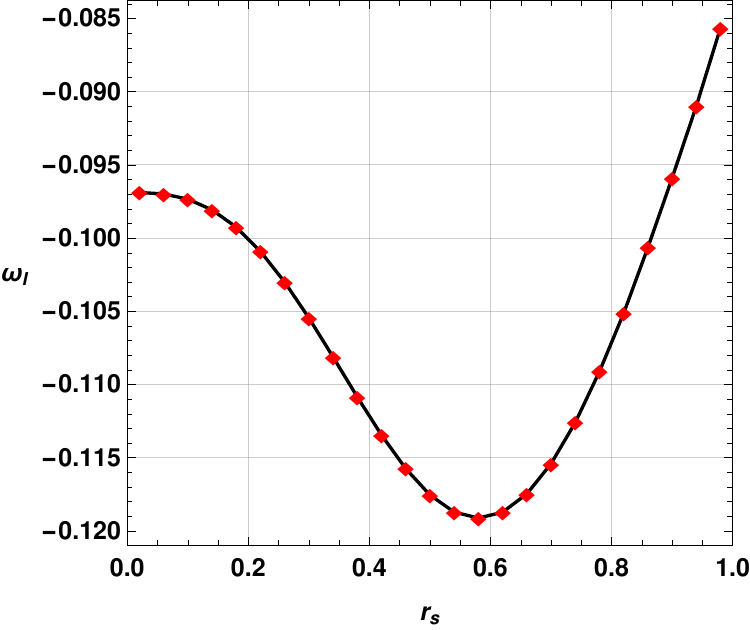}} \vspace{-0.2cm}
\caption{Variation of QNMs with respect to model parameter $r_s$
with $M=1$, $n= 0$, $\rho_s =0.5$, $a = 0.9$, $m_l=1$ and $l=2$ for scalar
perturbation.}
\label{QNMs02}
\end{figure}

The variation of QNM frequencies with respect to the Dehnen dark matter halo parameters-namely the central density $\rho_s$ and the halo radius $r_s$-provides important insights into how the galactic environment modifies the dynamics of a rotating BH. This analysis focuses on scalar perturbations with fixed BH parameters: mass $M = 1$, spin parameter $a = 0.9$, and quantum numbers $l = 2$, $m_l = 1$, and overtone number $n = 0$.

From Fig. \ref{QNMs01}, the left panel shows that the real part of the QNM frequency, $\omega_R$, decreases monotonically with increasing $\rho_s$. This indicates that higher central densities tend to lower the oscillation frequency of perturbations. Physically, this occurs because a denser halo contributes additional gravitational potential, which effectively deepens the trapping region for scalar waves, thus reducing their natural frequency of oscillation.
The imaginary part of the QNM frequency, $\omega_I$, is depicted in the right panel of Fig. \ref{QNMs01}. It shows a non-monotonic behavior with respect to $\rho_s$.  At first, it is seen that the absolute value $|\omega_I|$ rises in with $\rho_s$, which suggests a quicker damping or decay of the perturbation.  However, the damping begins to decrease after a certain value of $\rho_s$ (i.e at approximately 0.2).  This implies a transition: in less dense halos, a greater energy dissipation results from an increase in $\rho_s$, which enhances gravitational redshift occurring closer to the horizon.  However, the wave trapping becomes more important in denser environments, that results in less damped, longer-lived modes.

Now, by setting $\rho_s$ fixed at 0.5, we can observe from Fig. \ref{QNMs02} that displays comparable patterns with respect to $r_s$. As $r_s$ rises in the left panel, $\omega_R$ steadily falls. This is in line with the idea that larger halos soften the effective potential and reduce the oscillation frequency by extending the gravitational influence over a larger area.
Again, a non-monotonic profile is shown in the imaginary sector, which is displayed in the right panel of Fig. \ref{QNMs02}. Up to a certain value of $r_s$ (around 0.4), the damping rate rises (i.e., $|\omega_I|$ grows); after that, it begins to fall. This behavior implies that curvature gradients that improve scalar wave absorption are first strengthened by increasing $r_s$. But as the halo becomes more diffuse and extended, the effective potential tends to flatten, that reduces the wave dissipation and thereby allowing the perturbation to persist longer.

The aforementioned results suggest that both the real and imaginary parts of QNM frequencies are indeed sensitive to the presence and structure of the surrounding dark matter halo. The monotonic decrease in $\omega_R$ with increasing $\rho_s$ or $r_s$ implies that the natural oscillation frequencies of BHs embedded in dense or extended halos is lower than that in isolation. This effect may affect the spectral profile of gravitational waves emitted during the ringdown phase of BH mergers.
Moreover, the non-monotonic behavior of $\omega_I$ suggests that damping rates of gravitational wave signals can either increase or decrease depending on the specific dark matter configuration. Particularly, the presence of a local minima and maxima in the damping rate implies that the halo characteristics could be imprinted in the gravitational wave signal duration. This could possibly open a novel observational window into the study of galactic dark matter distributions via precision gravitational wave astronomy. Thus, these results emphasize the importance of accounting for environmental effects when modeling and interpreting BH dynamics and gravitational wave signals from both theory and observational perspectives. 

\section{Conclusion}
\label{conc}

In physically realistic scenarios, the BHs at the centres of galaxies are rotating. Therefore, in this work we have thoroughly investigated the optical and dynamical properties of a rotating BH embedded in a Dehnen‐type (1,4,0) galactic dark matter halo. Increasingly strong observational evidence for the presence of massive BHs in dwarf galaxies and galactic centres, as well as the ubiquity of galactic halos, motivates studying configurations in which a supermassive or intermediate‐mass BH resides within a core‐like dark matter halo characterised by a double power‐law density profile. Our methodology integrates cutting‐edge techniques in BH spacetime construction and analysis with recent theoretical advancements in dark matter profile modelling, offering insights into both fundamental physics and astrophysically realistic systems that could be probed by present and future observations.

This work extends our earlier results for a static Schwarzschild BH in a Dehnen halo, where the BH's gravitational potential is significantly altered by the density profile of the halo.  Importantly, we used the modified Newman–Janis method, developed by Azreg-A\"inou and others, to extend the non-rotating solution to incorporate rotation.  Using this approach, we were able to establish an explicit, physically consistent axisymmetric (Kerr-like) metric in which the BH's shape seamlessly incorporates the dark matter halo parameters, specifically the halo radius $r_s$ and the core density $\rho_s$.
The initial part of the study focused on understanding how the Dehnen dark matter halo affects the ergoregion and the event horizon, two basic geometric aspects of the rotating BH.  The horizon structure was shown to be significantly affected by dark matter parameters and rotation.  These findings suggest that energy extraction mechanisms from the spinning energy of the BH, such as the Penrose process operating in the ergoregion, may be boosted or repressed depending on the galaxy environment.
Next, we examined how light and null geodesics travel in this modified spacetime. Using the Hamilton–Jacobi method, we derived the specific equations that describe photon paths. We paid special attention to the conditions for unstable circular photon orbits. These orbits are crucial for forming the BH shadow, which appears as a dark silhouette against the background. This shadow is formed by photon paths spiraling around the photon sphere before either falling into the BH or escaping into space. Our calculations show that the size, shape, and asymmetry of the shadow are all affected by the Dehnen profile parameters. In particular, increasing $\rho_s$ or $r_s$ usually causes the shadow radius to grow while slightly altering its distortion for varying spin parameter $a$. We have also discussed how the distortion parameter $\delta$ is affected by the model parameters of the BH. In the next section, we studied the impacts of the model parameters on the emission rate of the BH. We have found that with an increase in the model parameters associated with this BH, the emission rate always decreases, as a result the life-time of the BH increases.

The BH's response to perturbations was thoroughly examined in the later part of the paper, by analyzing the QNMs of scalar fields, applying the WKB approximation and the Teukolsky formalism. QNMs represent the typical ringdown oscillations and decay frequencies of BHs after a disturbance, such as a merger or accretion, and reveal important information about the underlying spacetime. In our model, both the real (oscillation) and imaginary (damping) parts of the QNM frequencies were significantly influenced by the dark matter halo properties. Notably, the frequency consistently declines as $\rho_s$ and $r_s$ increase, indicating that denser or larger halos reduce the effective natural oscillation rate of the system. The damping (decay) rates show complex behavior, initially rising to a peak before falling again as halo density or scale increases. This implies a rich variety of outcomes, where different halo environments could produce both quickly and slowly decaying gravitational wave signals during the BH ringdown phase. These dependencies present a unique opportunity to use gravitational wave spectroscopy, especially with future detectors of greater sensitivity, to explore not just the BH parameters but also the otherwise hard-to-detect properties of nearby galactic dark matter halos.

Our results show that the relationship between a rotating BH and its surrounding galactic dark matter is crucial for both basic and observational astrophysics. Unlike simplified studies of BHs in empty space or basic environments, a Dehnen-type dark matter halo creates both new quantitative and qualitative effects in BH optics and dynamics.
The overall structure and methods used here can be developed further. For instance, while we have focused on the Dehnen (1,4,0) profile, the double power-law model includes a broader range of scenarios. This encompasses cored, cuspy, and triaxial halos that relate to different types of galaxies and cosmological situations. We can also apply these techniques to charged BHs, various theories of gravity, or situations with changing halos. Future directions are rich and necessary. Numerical relativity simulations that include non-vacuum backgrounds, studies of accretion disks and electromagnetic emission in these spacetimes, explicit calculations of strong-lensing signatures, and closer comparisons with observational data from EHT and gravitational wave observatories like LIGO-Virgo-KAGRA or the upcoming Einstein Telescope and LISA are essential. Additionally, considering the back-reaction of the BH on the halo, including baryonic matter, and exploring feedback mechanisms between jets and surrounding environments represent important next steps.

In summary, the results of our work suggests important implications of rotating BHs in realistic galactic environments. Our findings clearly show that features such as the BH shadow and QNM spectra, often seen as clean tests of rotating BH spacetimes, are sensitive to the properties of the surrounding dark matter halo. Far from being just theoretical oddities, these phenomena may soon be accessible through multi-messenger astronomy, allowing us to use BHs as unique tools for understanding both fundamental physics and the dark matter structure of the Universe.


\section*{Acknowledgment}
 DJG acknowledges the contribution of the COST Action CA21136  -- ``Addressing observational tensions in cosmology with systematics and fundamental physics (CosmoVerse)". 

\section*{Declaration of competing interest}
The authors declare that they have no known competing financial interests or personal relationships that could have appeared to influence the work reported in this manuscript.

\section*{Data Availability Statement}
There are no new data associated with this article.

\bibliographystyle{apsrev4-2}
\bibliography{biblio.bib}

\end{document}